\documentclass[lettersize,journal]{IEEEtran}
\usepackage{amsmath,amsfonts}
\usepackage{algorithmic}
\usepackage{algorithm}
\usepackage{array}
\usepackage[caption=false,font=normalsize,labelfon
t=sf,textfont=sf]{subfig}
\usepackage{textcomp}
\usepackage{stfloats}
\usepackage{url}
\usepackage{verbatim}
\usepackage{graphicx}
\usepackage{cite}
\usepackage{xcolor}
\usepackage{bbding}
\usepackage{multirow}
\usepackage{booktabs}
\usepackage{makecell}
\usepackage{changepage}

\title{
\textit{EidetiCom}: A Cross-modal Brain-Computer Semantic Communication Paradigm for Decoding Visual Perception
}

\author{Authors}

\author{Linfeng~Zheng,
        Peilin~Chen, 
        and Shiqi~Wang,~\textit{Senior Member,~IEEE}% <-this % stops a space

\thanks{Linfeng Zheng, Peilin Chen, and Shiqi Wang are with the Department of Computer Science, City University of Hong Kong, Hong Kong, China. (e-mail: lfzheng3-c@my.cityu.edu.hk, \{plchen3, shiqwang\}@cityu.edu.hk).} 

}

\begin{document}
% \linenumbers

\maketitle

\begin{abstract}
Brain-computer interface (BCI) facilitates direct communication between the human brain and external systems by utilizing brain signals, eliminating the need for conventional communication methods such as speaking, writing, or typing. Nevertheless, the continuous generation of brain signals in BCI frameworks poses challenges for efficient storage and real-time transmission. While considering the human brain as a semantic source, the meaningful information associated with cognitive activities often gets obscured by substantial noise present in acquired brain signals, resulting in abundant redundancy. In this paper, we propose a cross-modal brain-computer semantic communication paradigm, named \textit{EidetiCom}, for decoding visual perception under limited-bandwidth constraint. The framework consists of three hierarchical layers, each responsible for compressing the semantic information of brain signals into representative features. These low-dimensional compact features are transmitted and converted into semantically meaningful representations at the receiver side,  serving three distinct tasks for decoding visual perception: brain signal-based visual classification, brain-to-caption translation, and brain-to-image generation, in a scalable manner. Through extensive qualitative and quantitative experiments, we demonstrate that the proposed paradigm facilitates the semantic communication under low bit rate conditions ranging from 0.017 to 0.192 bits-per-sample, achieving high-quality semantic reconstruction and highlighting its potential for efficient storage and real-time communication of brain recordings in BCI applications, such as eidetic memory storage and assistive communication for patients.
\end{abstract}


\section{Introduction}

Brain-computer interface (BCI) is an emerging technology that intersects neuroscience, cognitive science, signal processing, and artificial intelligence. This technology establishes a direct communication pathway between the human brain and a computer, enabling individuals to interact with external systems without relying on traditional forms of communication like speaking, writing, or typing \cite{chaudhary2016brain, schultz2017biosignal, willett2021high}. BCI holds significant value for researchers as it facilitates the study of neural activities of the human brain, leading to a better understanding of its functionalities and mechanisms \cite{halder2011neural}. Moreover, BCI opens up diverse applications. For instance, by analyzing the intention behind brain activity, it becomes possible to decode brain signals and utilize them to control external devices such as robotic prosthetics \cite{edelman2019noninvasive} or digital avatars \cite{metzger2023high}. Additionally, BCI offers a more convenient means of exploring cognitive processes and human awareness, including mental and psychological analysis \cite{roc2021review}, analysis of affective states \cite{alarcao2017emotions}, and even neural decoding of semantic concepts like text \cite{willett2021high}, speech \cite{metzger2023high,defossez2023decoding}, or visual imagery \cite{horikawa2017generic,wang2023better}.

Recently, advancements in representation learning and generative models have greatly facilitated the neural decoding of brain signals, making it increasingly feasible and practical. Among these research efforts, the neural decoding of visually-evoked brain signals has received considerable attention. Typically, non-invasive technologies such as functional magnetic resonance imaging (fMRI), electroencephalography (EEG), and magnetoencephalography (MEG) are used to capture the brain signals related to visual stimuli. Most related works focus on distinguishing the category of visual stimuli \cite{spampinato2017deep,lawhern2018eegnet,tao2021gated} or reconstructing visual stimuli \cite{kavasidis2017brain2image,tirupattur2018thoughtviz,takagi2023high,bai2023dreamdiffusion,chen2023seeing} shown to the experimental subjects. For visual classification, multimodal approaches \cite{palazzo2020decoding,du2023decoding} have proven effective in improving classification accuracy. For visual reconstruction, the usage of generation models, such as variational autoencoder (VAE) \cite{spampinato2017deep}, generative adversarial network (GAN) \cite{spampinato2017deep,tirupattur2018thoughtviz}, and latent diffusion models \cite{takagi2023high,lan2023seeing}, have significantly improved the realism of the reconstructed visual stimuli.

However, the raw brain signals acquired by BCI frameworks demand extensive bandwidth and storage capacity, posing challenges for both efficient storage and real-time transmission. For instance, when we consider a 128-channel EEG-based brain-computer interface operating at a sampling rate of 1000 Hz and with a data precision of 16 bits, it would generate 2.048 megabits of data per second. For missions such as eidetic memory or dream storage, which require the recording of visually-evoked cognitive activities over extended periods, storing all raw signals would impose significant storage burdens on users. Similarly, in missions like assistive communication for patients, where the cognitive activities of patients need to be transmitted in real-time as auxiliary information to remote healthcare providers or family members, the limited bandwidth in real-world scenarios makes it challenging to support the real-time transmission of raw signals. To reduce the storage and transmission cost for downstream applications, it is feasible to utilize existing scientific data compression methods, such as SZ3 \cite{liang2022sz3} and ZFP \cite{lindstrom2014fixed}, which have been developed to reduce the data size of floating-point data. However, they are not specifically developed for brain signals, ignoring how the brain signals are finally utilized. Therefore, there is an increasing demand to develop a brain signal communication framework that features high flexibility, low bandwidth consumption, and cross-modal representation, supporting diverse BCI applications in narrow-band communication scenarios.

\begin{figure*}
    \centering
    \includegraphics[width=0.77\linewidth]{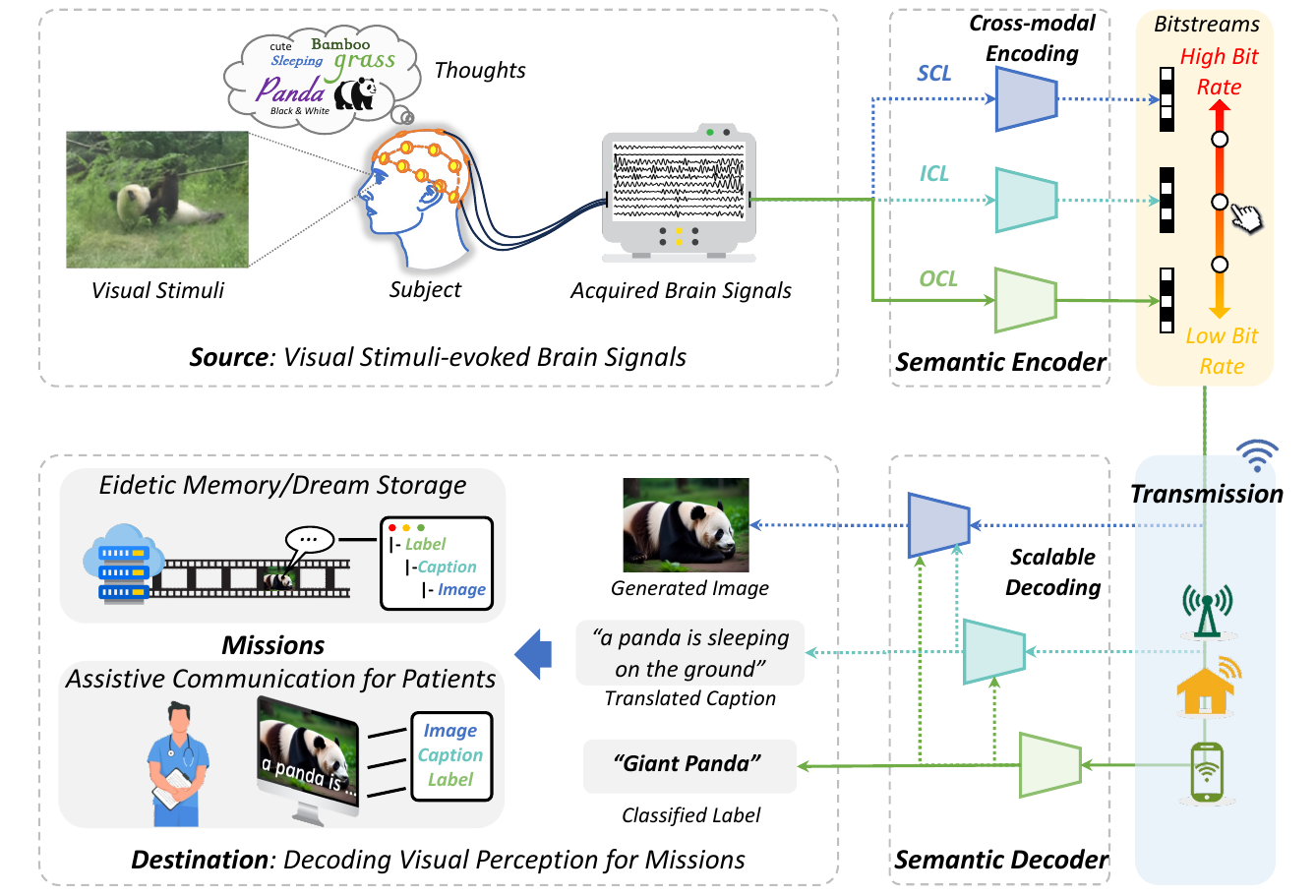}
    \caption{Overall framework of \textit{EidetiCom}, a cross-modal brain-computer semantic communication paradigm for decoding visual perception. The \textit{EidetiCom} framework encapsulates the entire information processing chain, incorporating the information source, semantic encoder, transmission link, semantic decoder, and information destination. In this semantic communication paradigm, the human brain serves as the source, generating visually-evoked brain signals associated with cognition activities. These signals undergo encoding, transmission, and decoding and are ultimately absorbed by the destination, which decodes visual perception for missions. The \textit{EidetiCom} framework significantly reduces the storage and transmission costs of brain signals for specific missions, such as eidetic memory or dream storage, which are sensitive to storage needs, and assistive communication for patients, which is sensitive to bandwidth requirements.}
    \label{fig:overall}
\end{figure*}

We address these grand challenges from a new perspective, as brain signals are intrinsically utilized for particular missions. Therefore, we propose a mission-oriented brain-computer semantic communication paradigm \textit{EidetiCom}, which is the first attempt to make the brain signals compact based on the ultimate mission. Herein, we focus on the most challenging mission - decoding visual perception, including three tasks: brain signal-based visual classification, brain-to-caption translation, and brain-to-image generation. In this work, an innovative cross-modal brain-computer semantic communication paradigm is developed. It is specifically designed to extract and compress the semantic information of visually-evoked brain signals for decoding visual perception, as semantic information is mostly concerned in neural decoding tasks while low-level structure information of visual stimuli are often lost in the brain signals. Therefore, we can effectively eliminate irrelevant information in the brain signals, preserving the semantic-relevant data at a much lower bit rate.

The scalable coding framework of our method is structured hierarchically, capitalizing on the strong semantic connections among the three tasks. As illustrated in Fig.~\ref{fig:overall}, 
The proposed framework is composed of three layers, including the Object-level Category Layer (OCL) for visual classification, the Image-level Caption Layer (ICL) for inferring caption, and the Stimuli-level Cognition Layer (SCL) for reconstructing a thumbnail of the visual stimuli. The scalable representation enables the reconstruction of the higher layer to benefit from the outputs of lower layers, further guaranteeing the effectiveness of communication. Moreover, this systematic design allows for precise and adaptable control of the transmission bit rate, meeting the distinct demands of the missions involved. By leveraging these layers, we can effectively compress the brain signals and recover the essential semantic information that can be used for various downstream applications.

As the first paradigm that achieves cross-modal semantic communication of brain signals in a scalable way, 
the proposed \textit{EidetiCom} advances brain-computer communication from various perspectives. The main contributions of this work can be summarized as follows. First, we achieve $83\times\sim941\times$ compression ratio of brain signals without sacrificing the performance of the final task, showing great potential for real-time communication of brain activities. Second, \textit{EidetiCom} facilitates the understanding of brain activities based on a scalable pipeline that supports three distinct tasks for decoding visual perception: brain signal-based visual classification, brain-to-caption translation, and brain-to-image generation. 
By utilizing the correlations between these tasks, our pipeline can support efficient and scalable decoding for different visual perception goals.
Finally, the proposed mission-oriented communication paradigm, which supports multiple missions simultaneously, can shed light on a wide variety of applications beyond BCI.

\section{Related Works}

\subsection{Neural Decoding of Brain Signal for BCI}
Neural decoding of brain signals is a fundamental approach that enables researchers to uncover the underlying neural activities of the human brain. In the field of brain research, various types of non-invasive brain signals are commonly utilized, such as functional magnetic resonance imaging (fMRI), electroencephalography (EEG), and magnetoencephalography (MEG). Each type of brain signal possesses unique characteristics. For instance, EEG signals acquired through non-invasive brain-computer interfaces are considered safe and portable, offering high temporal resolution compared to invasive signals. With the increasing availability of brain signal datasets \cite{hanke2014high,gifford2022large}, research on brain signal decoding has become more practical. In recent literature, there have been notable advancements in reconstructing stimuli, such as audio and images from brain signals. Metzger \textit{et al.} \cite{metzger2023high} utilized high-density cortical recordings from a participant with paralysis to enable real-time, high-fidelity speech decoding in text, audio, and facial-avatar animation modalities. D{\'e}fossez \textit{et al.} \cite{defossez2023decoding} proposed a deep model trained via contrastive learning to decode self-supervised representations of perceived speech from MEG recordings. Chen \textit{et al.} \cite{chen2024neural} proposed a deep learning-based neural speech decoding framework with an ECoG-to-speech decoder and a differentiable speech synthesizer. Scotti \textit{et al.} \cite{scotti2024reconstructing} proposed to retrieve and reconstruct viewed images from fMRI signals. Kavasidis \textit{et al.} \cite{kavasidis2017brain2image} proposed a generative adversarial network-based network to reconstruct images from EEG signals. Ren \textit{et al.} \cite{ren2021reconstructing} proposed a method that maps the raw EEG signals to a visual feature space and subsequently decodes the features into an image. Lan \textit{et al.} \cite{lan2023seeing} introduced a latent diffusion model to generate high-quality semantic reconstructions from EEG signals. 

Considering the practical value of EEG signals in brain research and their potential for non-invasive applications, it is crucial to explore methods for compressing and transmitting EEG signals effectively. This motivates our research to develop novel compact representation techniques specifically tailored for EEG signal compression. We pay special attention to the ultimate utilization of these brain signals, ensuring that our techniques align with their intended applications.

\subsection{End-to-end Signal Compression}
In recent years, numerous successes have been witnessed in learning-based end-to-end signal compression, especially end-to-end image and video compression. Ballé \textit{et al.} proposed the end-to-end image compression framework \cite{balle2016end} based on autoencoder with a fully factorized entropy model and further extended this framework with hyperprior \cite{balle2018variational} and autoregressive model \cite{minnen2018joint} for accurate entropy estimation. 
Nowadays, the most advanced end-to-end image compression methods have surpassed traditional methods in terms of rate-distortion performance~\cite{liu2023learned}.
In addition to images, researchers conducted similar approaches to compress various modalities of signals, such as videos \cite{lu2019dvc}, point clouds \cite{wang2021multiscale}, genetic sequences \cite{wang2018deepdna}, speech and audio signals \cite{zeghidour2021soundstream}. 
Generally speaking, given an input signal $\boldsymbol{x}$, the encoder transforms the signal into a compact latent feature $\boldsymbol{y}$, and the decoder recovers the quantized latent feature $\boldsymbol{\hat{y}}$ into reconstructed signal $\boldsymbol{\hat{x}}$. The optimization objective function of end-to-end compression is given by,
\begin{equation}
\begin{aligned}
J &= R(\boldsymbol{\hat{y}}) + \lambda D(\boldsymbol{x}, \boldsymbol{\hat{x}}) \\ &= \mathbb{E}_{\boldsymbol{x} \sim p_{\boldsymbol{x}} } [{-\log p_{\boldsymbol{\hat{y}}}}(\boldsymbol{\hat{y}})]  + \lambda \mathbb{E}_{\boldsymbol{x} \sim p_{\boldsymbol{x}}} [d(\boldsymbol{x},\boldsymbol{\hat{x}})], 
\end{aligned}
\end{equation}
where $R$ denotes the bit rate of quantized feature $\boldsymbol{\hat{y}}$, $D$ denotes the distortion, $p_{\boldsymbol{x}}$ and $p_{\boldsymbol{\hat{y}}}$ denotes the probability distribution of $\boldsymbol{x}$ and $\boldsymbol{\hat{y}}$ respectively, $d(\cdot, \cdot)$ measures the distortion between original signal $\boldsymbol{x}$ and reconstructed signal $\boldsymbol{\hat{x}}$, and $\lambda$ is the Lagrange multiplier that controls the trade-off between bit rate $R$ and distortion $D$.

In addition to signal reconstruction, the received reconstructed signal or feature can also be used to perform various downstream tasks. This compression paradigm is called task-oriented compression \cite{yang2023introduction}. In this case, the distortion term in the objective function may be replaced by a specific task loss, such as the cross-entropy loss for classification, or a contrastive loss function for self-supervised learning \cite{dubois2021lossy} to preserve generic semantic information. To enhance flexibility, scalable coding pipelines \cite{yan2021sssic, zhang2023rethinking, mao2023scalable} have been developed to support multiple tasks.

\subsection{Cross-modal Semantic Communication}
The core concept of semantic communication originally stems from Shannon's seminal work \cite{shannon2001mathematical}. Based on this concept, Weaver further categorized communication into three levels: symbol level, semantic level, and effective level \cite{weaver1953recent}. 
In recent years, substantial attention has been paid to semantic communication due to its potential in ultra-low bit rate data communication. 
In a semantic communication paradigm \cite{gunduz2022beyond}, the semantic information from the source signals is extracted and transmitted through the link, while irrelevant and redundant information is filtered out. 
Based on this principle, Xie \textit{et al.} \cite{xie2022task} proposed a transformer-based semantic communication framework for text transmission, aiming to preserve the semantic information of sentences.
To achieve a more accurate evaluation of textual data's semantic fidelity, Wang \textit{et al.} \cite{wang2022performance} introduced a metric of semantic similarity based on semantic accuracy and completeness for text transmission. 
For image transmission, Huang \textit{et al.} \cite{huang2022toward} propose a reinforcement learning-based adaptive coding approach to encode visual semantic information. 
When the reconstructed signal is primarily applied to downstream tasks, the paradigm is referred to as task-oriented communication. 
For scene classification tasks, Kang \textit{et al.} \cite{kang2022task} proposed an aerial image transmission scheme to achieve the trade-off between transmission delay and classification accuracy.
Xie \textit{et al.} \cite{xie2022task} proposed a transformer-based framework for transmitting multimodal data in multi-user scenarios, supporting intelligent tasks such as image retrieval, machine translation, and visual question answering.

\begin{figure*}
    \centering
    \includegraphics[width=0.65\linewidth]{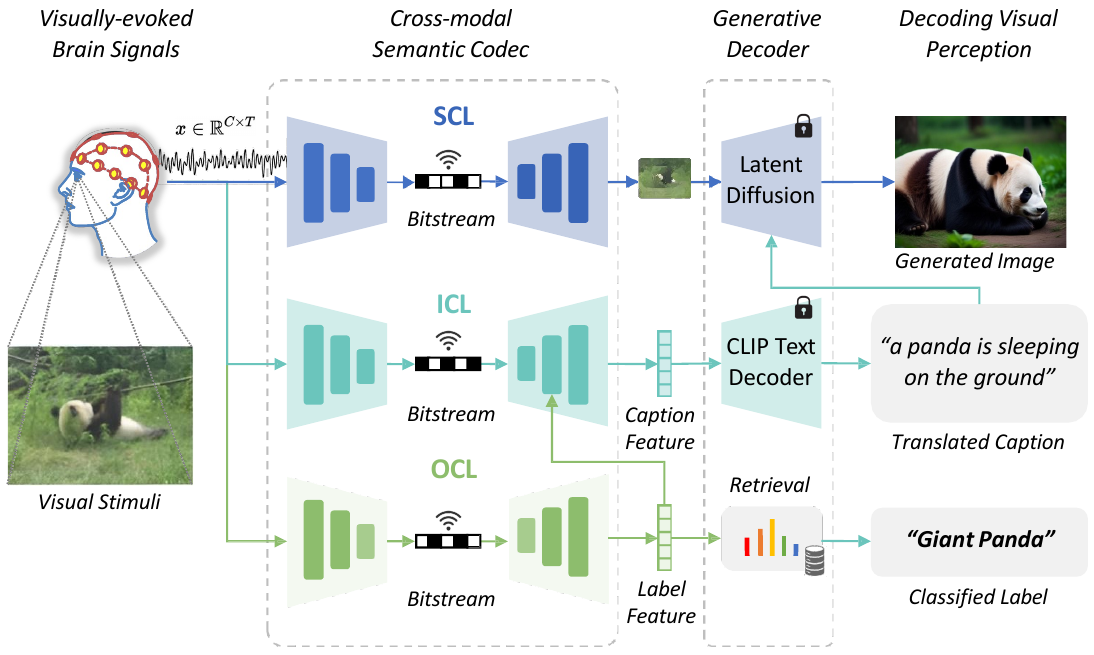}
    \caption{The three-layered codec architecture of \textit{EidetiCom}. \textit{EidetiCom} consists of three hierarchical layers, each responsible for compressing visually-evoked brain signals into compact semantic features, facilitating effective transmission. These features can be utilized in a scalable manner to decode the visual category, image caption, and generated image associated with the visual stimuli.}
    \label{fig:overall}
\end{figure*}

\section{Method}

\subsection{The Overall Framework}
The overall framework of our innovative cross-modal brain-computer semantic communication paradigm for decoding visual perception is illustrated in Fig. \ref{fig:overall}. 
\textit{EidetiCom} consists of three layers, each responsible for compressing visually-evoked brain signals into compact semantic features, facilitating effective transmission. 
These features can be utilized in a scalable manner to decode the visual category, image caption, and generated image associated with the visual stimuli. 
The proposed three-layered coding framework showcases its capability to effectively support three distinct semantic-relevant tasks in a scalable manner: brain signal-based visual classification, brain-to-caption translation, and brain-to-image generation.

For a specific mission, such as assistive communication for patients, the granularity of the decoded visual perception depends on user requirements and the real-world constraints, such as transmission bandwidth limitations. For instance, during a routine inspection, healthcare providers may only need to examine the label of the patient's visual perception. However, in special circumstances requiring more detailed information, it may be necessary to decode the patient's data into captions or images with richer semantic content. Additionally, fluctuations in transmission bandwidth can make it difficult to effectively transmit high bit rate contents. Therefore, designing a scalable encoding framework helps to flexibly adjust the decoded content and transmission bit rate under varying user requirements and environmental conditions. Specifically, we design a three-layered scalable coding framework, including OCL, ICL and SCL, corresponding to three distinct levels of semantic information in visual perception: label, caption, and image. Each level contains semantic information of varying granularity, enabling users to select the most appropriate level based on specific requirements and transmission conditions.

Due to the semantic correlation between layers, encoding and decoding features independently would lead to redundancy. To address this, we employ conditional decoding, leveraging semantic information from higher levels as side information for decoding lower-level semantic information. Specifically, information at the label level serves as a condition for decoding both caption and image information, while caption-level information conditions the decoding of image information. Typically, given a visual stimuli–evoked EEG signal $\boldsymbol{x} \in \mathbb{R}^{C \times T}$, where $C$ and $T$ represent the channel number and time period of brain signal respectively, the end-to-end learnable codecs of the three layers transform the signal into separate label, caption and thumbnail features. The label feature is used for classification through retrieval. The caption feature, decoded with side information from the label feature, is used for generating captions. The reconstructed thumbnail, is used for generating image with the side information provided by label and caption information. This approach allows for a flexible selection of decoded content and bit rate according to changes in mission requirements and transmission conditions.

\subsection{Object-level Category Layer}
The OCL of the framework is responsible for classifying brain signals based on brain-to-label semantic coding. Brain signal classification involves distinguishing the category of visual stimuli. Typically, when given visual stimuli–evoked EEG signal $\boldsymbol{x}$, a classifier $f$ maps the original signal to the normalized probability $f(\boldsymbol{x}) \in [0,1]^{N}$ of $N$ categories. The classifier is usually optimized according to cross-entropy loss between predicted probability and ground truth one-hot labels.

\begin{figure}
    % \centering
    \includegraphics[width=0.95\linewidth]{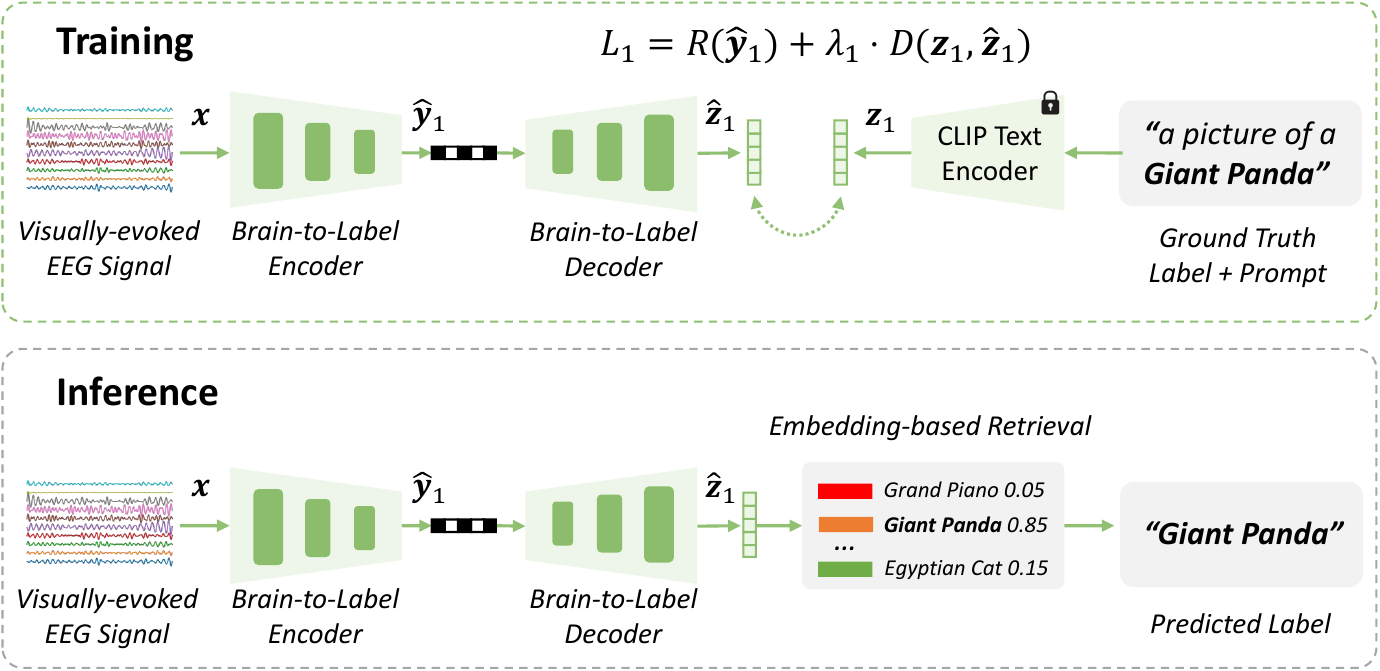}
    \caption{The architecture of Object-level Category Layer (OCL). During training, the layer is optimized to achieve the best trade-off between bit rate and the distortion of semantic information for the label. During inference, the compressed features are utilized to identify the category of visually-evoked brain signals based on cosine similarities.}
    \label{fig:Layer1}
\end{figure}

In order to align the semantic content within brain signals with label embeddings, we introduce a brain-to-label semantic coding strategy that leverages the decoded semantic features for executing a retrieval-based classification. 
Firstly, we leverage a pretrained CLIP model \cite{radford2021learning} to extract the semantic features $\{\boldsymbol{F}_1, \boldsymbol{F}_2, ..., \boldsymbol{F}_N\}$ from the texts of ground truth labels $\{\boldsymbol{l}_1, \boldsymbol{l}_2, ..., \boldsymbol{l}_N\}$ and store these semantic features of labels in a database. 
Secondly, a learned semantic feature extractor will map the brain signal $\boldsymbol{x}$ to the corresponding semantic feature $\boldsymbol{F_x}$. 
The cosine similarity between $\boldsymbol{F_x}$ and each feature from the database will be calculated, and the category $n$ with the highest similarity score will be selected as the predicted class $\boldsymbol{l}_{n^{*}}$, i.e.,
\begin{equation}
\quad n^{*} = \operatorname*{argmax}_{n \in \{ 1, 2, ..., N\}} \cos (\boldsymbol{F_x}, \boldsymbol{F}_n),
\end{equation}
where $\cos(\cdot, \cdot)$ is the cosine similarity.

To enable efficient storage or transmission, we incorporate a bottleneck within the semantic feature extractor to extract compact latent representation. 
This partitions the cross-modal feature extractor into an encoder and a decoder. As shown in Fig. \ref{fig:Layer1}, the transform coding process in the OCL can be described as follows:
\begin{equation}
\begin{aligned}
\boldsymbol{y}_1 &= f_{\text{enc}}^1 (\boldsymbol{x} ; \boldsymbol{\theta}_1) \\
\boldsymbol{\hat{y}}_1 &= Q (\boldsymbol{y}_1) \\
\boldsymbol{\hat{z}}_1 &= f_{\text{dec}}^1 (\boldsymbol{\hat{y}}_1 ; \boldsymbol{\phi}_1).
\end{aligned}
\end{equation}
The encoder $f_{\text{enc}}^1(\cdot ; \boldsymbol{\theta}_1)$ maps the raw brain signal $\boldsymbol{x}$ to a temporal aggregated latent feature $\boldsymbol{y}_1 \in \mathbb{R}^{C_1}$. The feature $\boldsymbol{y}_1$ is then quantized by a rounding function $Q(\cdot)$, yielding a quantized feature $\boldsymbol{\hat{y}}_1$. Subsequently, a fully factorized \cite{balle2018variational} entropy model $p(\boldsymbol{\hat{y}}) = \prod_{i} p(\boldsymbol{\hat{y}}^i)$, where $\boldsymbol{\hat{y}}^i$ denotes the $i$-th channel of $\boldsymbol{\hat{y}}$, estimates the probability density function (PDF) of the latent feature. 
The latent feature can be losslessly compressed to binary bitstreams based on the estimated PDF using an entropy coding method, e.g., arithmetic coding. 
Upon receiving $\boldsymbol{\hat{y}}_1$, the decoder $f_{\text{dec}}^1(\cdot;\boldsymbol{\phi}_1)$ transforms the latent feature into a text-level semantic feature $\boldsymbol{\hat{z}}_1 \in \mathbb{R}^{C_{1}^{'}}$.

During training, the non-differentiable hard quantization is replaced with additive uniform noise for joint optimization, i.e., $\boldsymbol{\hat{y}_1}=\boldsymbol{y}_1 + \boldsymbol{u}, \boldsymbol{u} \sim \mathcal{U}(-0.5, 0.5)$. The codec will be optimized to achieve the best trade-off between bit rate and brain-to-label cross-modal feature distortion based on the specified rate-distortion loss function:
\begin{equation}
\begin{aligned}
\mathcal L_1 &= R(\boldsymbol{\hat{y}}_1) + \lambda_1 D(\boldsymbol{z}_1, \boldsymbol{\hat{z}}_1) \\ &= \mathbb{E}_{\boldsymbol{x} \sim p_{\boldsymbol{x}}} [{-\log p_{\boldsymbol{\hat{y}}_1}}(\boldsymbol{\hat{y}}_1)]  + \lambda_1 \mathbb{E}_{\boldsymbol{x} \sim p_{\boldsymbol{x}}} [d_1(\boldsymbol{z}_1,\boldsymbol{\hat{z}}_1)],
\end{aligned}
\end{equation}
where $\boldsymbol{z}_1=\text{CLIP}_{\textit{text}}(\boldsymbol{t}_{\textit{label}})$, and $\boldsymbol{t}_{\textit{label}}$ is the label of the brain signal $\boldsymbol{x}$ embedded within a natural language prompt. 
The loss function aims to minimize the difference between the semantic feature of the ground truth label and the semantic feature predicted from the brain signal while maintaining superior compression performance. 
The distortion measure $d_1(\cdot,\cdot)$ is defined as the combination of mean squared error (MSE) and cosine similarity, i.e., 
\begin{equation}
d_1 (\boldsymbol{z}_1, \boldsymbol{\hat{z}}_1) =  \textit{MSE}( \boldsymbol{z}_1,  \boldsymbol{\hat{z}}_1) + \alpha_1 \cdot ( 1 - \cos(\boldsymbol{z}_1, \boldsymbol{\hat{z}_1}) ),
\end{equation}
where $\alpha_1$ controls the balance of the two terms. 

During inference, the predicted label, whose corresponding semantic feature has the highest similarity score with the received semantic feature $\hat{y}_1$, will be retrieved from the pre-built database.

\begin{figure}
    \centering
    \includegraphics[width=0.95\linewidth]{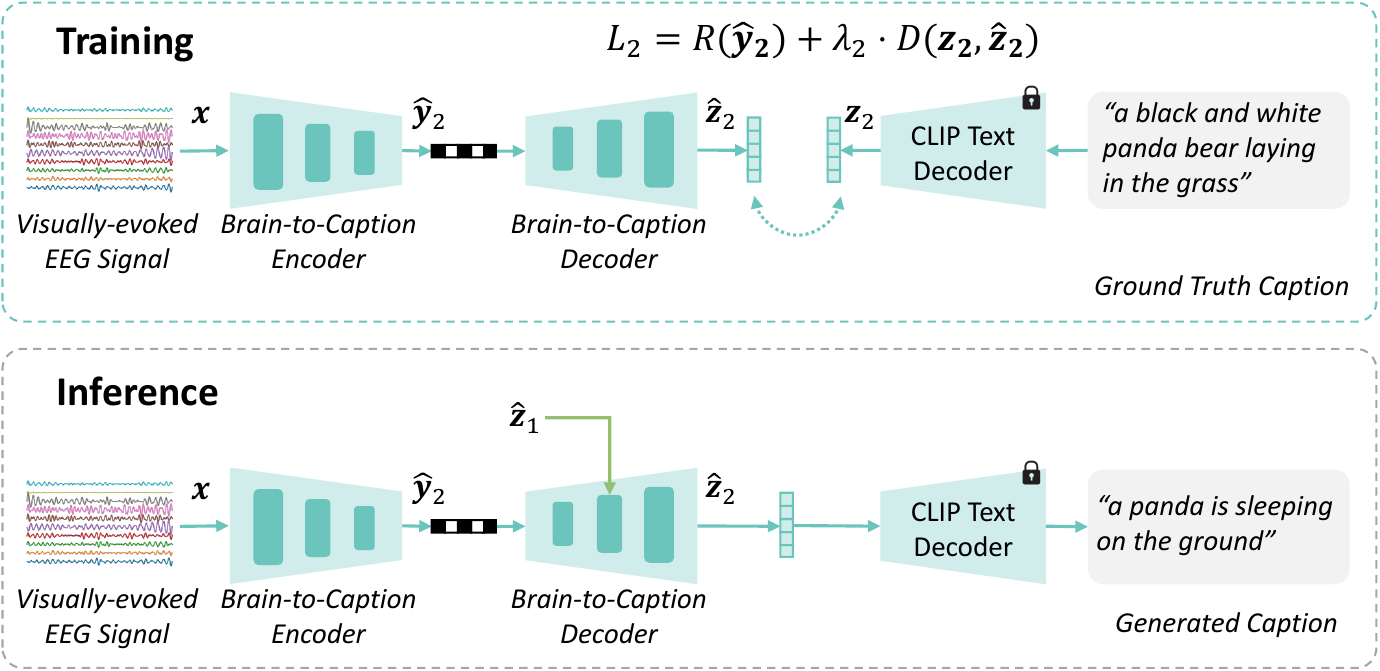}
    \caption{The architecture of Image-level Caption Layer (ICL). During training, the layer is optimized to achieve the best trade-off between the bit rate and the distortion of semantic information for the caption. During inference, the compressed features are translated to image captions for the visual stimuli using the pretrained CLIP text decoder.}
    \label{fig:Layer2}
\end{figure}

\subsection{Image-level Caption Layer}
The ICL is built upon the foundation of the OCL and is specifically dedicated to brain-to-caption translation. It's noteworthy that the OCL's decoder produces a coarse text-level semantic feature aligned with the label. 
This semantic feature can then serve as a natural condition for decoding a more detailed text-level semantic feature associated with a complete sentence. 
Consequently, this detailed feature can be seamlessly translated into easily understandable sentences for human comprehension.

\begin{figure}
    \centering
    \includegraphics[width=0.95\linewidth]{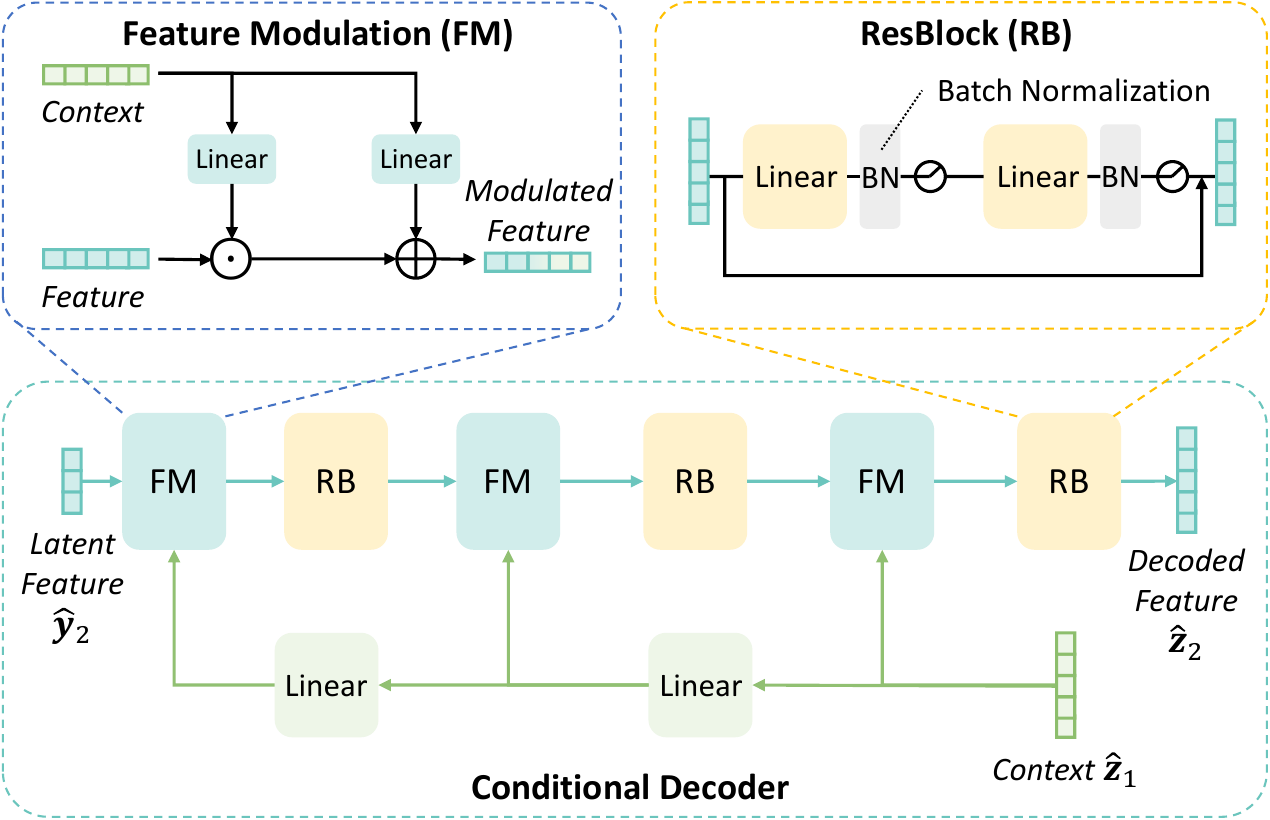}
    \caption{The conditional decoder of ICL. The latent feature $\boldsymbol{\hat{y}}_2$ is decoded into a more detailed text-level semantic feature $\boldsymbol{\hat{z}}_2$ with the modulation of category context $\boldsymbol{\hat{z}}_1$.}
    \label{fig:Layer2_Decoder}
\end{figure}

As shown in Fig. \ref{fig:Layer2}, utilizing the conditional decoding approach, the transform coding process in the ICL can be described as follows:
\begin{equation}
\begin{aligned}
\boldsymbol{y}_2 &= f_{\text{enc}}^2 (\boldsymbol{x} ; \boldsymbol{\theta}_2) \\
\boldsymbol{\hat{y}_2} &= Q (\boldsymbol{y}_2) \\
\boldsymbol{\hat{z}_2} &= f_{\text{dec}}^2 (\boldsymbol{\hat{y}_2} , \boldsymbol{\hat{z}_1} ; \boldsymbol{\phi}_2) .
\end{aligned}
\end{equation}
The encoder $f_{\text{enc}}^2(\cdot ; \boldsymbol{\theta}_2)$ maps the original brain signal $\boldsymbol{x}$ to a temporal aggregated latent feature $\boldsymbol{y}_2 \in \mathbb{R}^{C_2}$.
Similar to the OCL, the latent feature $\boldsymbol{y}_2$ is also rounded and yields the quantized feature $\boldsymbol{\hat{y}}_2 = Q(\boldsymbol{y}_2)$, and the fully factorized entropy model is used to estimate the PDF of the latent feature. 
Conditioned on the decoded semantic feature $\boldsymbol{\hat{z}}_1$ of the OCL, the decoder $f_{\text{dec}}^2(\cdot;\boldsymbol{\phi}_2)$ transforms the latent feature $\boldsymbol{\hat{y}}_2$ into a more detailed text-level semantic feature $\boldsymbol{\hat{z}}_2  \in \mathbb{R}^{C_{2}^{'}} $.
As illustrated in Fig. \ref{fig:Layer2_Decoder}, the conditional decoder is composed of several residual blocks and feature modulation modules in cascade. The feature modulation module $\text{FM}(\cdot, \cdot)$ transforms the semantic feature $\boldsymbol{\hat{z}_1}$ decoded from the OCL into affine transformation parameters $\boldsymbol{\gamma}$ and $\boldsymbol{\beta}$. These parameters are subsequently employed to scale and shift the hidden feature $\boldsymbol{h}$ in the decoder of the ICL, i.e.:
\begin{equation}
\text{FM}(\boldsymbol{h}, \boldsymbol{\hat{z}_1}) = \boldsymbol{\gamma} \odot \boldsymbol{h}  + \boldsymbol{\beta},
\end{equation}
where $\odot$ denotes element-wise multiplication.

During training, the rounding operation is replaced by additive uniform noise, and the codec is optimized to achieve the best trade-off between bit rate and brain-to-caption cross-modal feature distortion using the following rate-distortion loss function:
\begin{equation}
\begin{aligned}
\mathcal L_2 &= R(\boldsymbol{\hat{y}}_2) + \lambda_2 D(\boldsymbol{z}_2, \boldsymbol{\hat{z}}_2) \\ &= \mathbb{E}_{\boldsymbol{x} \sim p_{\boldsymbol{x}}} [{-\log p_{\boldsymbol{\hat{y}}_2}}(\boldsymbol{\hat{y}}_2)]  + \lambda_2 \mathbb{E}_{\boldsymbol{x} \sim p_{\boldsymbol{x}}} [d_2(\boldsymbol{z}_2,\boldsymbol{\hat{z}}_2)],
\end{aligned}
\end{equation}
where $\boldsymbol{z}_2 = \text{CLIP}_{\textit{text}}(\boldsymbol{t}_{\textit{caption}})$, and $\boldsymbol{t}_{\textit{caption}}$ is the image caption of visual stimuli corresponding to the brain signal $\boldsymbol{x}$. The loss function aims to minimize the discrepancy between the semantic feature of the ground truth image caption and the semantic feature predicted from the brain signal. Additionally, it constrains the bit rate of the compact feature. The distortion measure $d_2(\cdot, \cdot)$ is the combination of MSE and cosine similarity, i.e., 
\begin{equation}
d_2 (\boldsymbol{z}_2, \boldsymbol{\hat{z}}_2) =  \textit{MSE} (\boldsymbol{z}_2 , \boldsymbol{\hat{z}}_2) + \alpha_2 \cdot ( 1 - \cos(\boldsymbol{z}_2, \boldsymbol{\hat{z}}_2) ),
\end{equation}
where $\alpha_2$ controls the balance of the two terms. 

During inference, the semantic feature $\boldsymbol{\hat{z}_2}$ is used to generate a human-comprehensible caption associated with the visual stimuli based on the CLIP text decoder of the pretrained DeCap model \cite{lidecap}.

\begin{figure}
    \centering
    \includegraphics[width=0.95\linewidth]{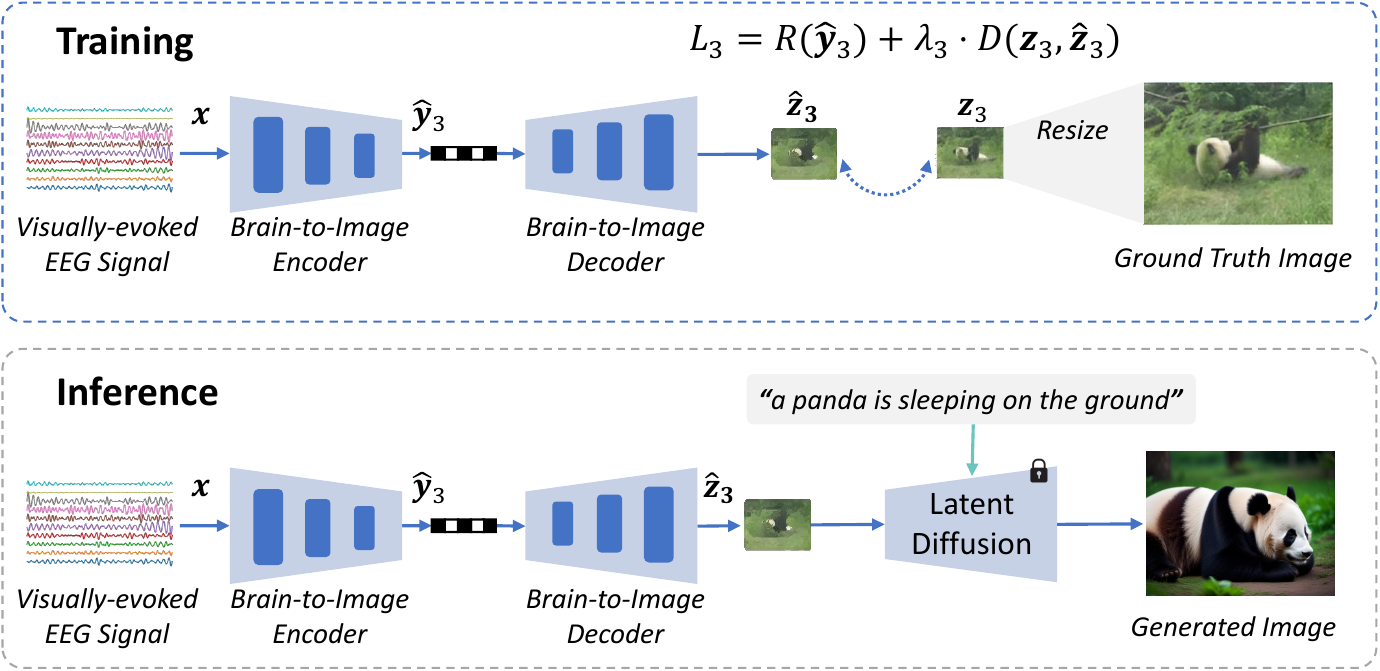}
    \caption{The architecture of Stimuli-level Cognition Layer (SCL). During training, the layer is optimized to achieve the best trade-off between the bit rate and the distortion of the reconstructed thumbnail. During inference, the thumbnail and translated caption are used to generate an image associated with the visual stimuli, based on a pretrained latent diffusion model.}
    \label{fig:Layer3}
\end{figure}

\subsection{Stimuli-level Cognition Layer}
The SCL focuses on image decoding based on conditional image generation using brain signals. Existing methods for conditional image generation are capable of generating realistic images based on specific text prompts. In this case, the output of the ICL, which is a caption associated with the visual stimuli, can be used to generate an image. Thus, the SCL only requires a conditional image generator and does not need any additional information to be encoded into bitstreams for transmission. 

However, the caption derived from the brain signal only contains high-level semantic information and lacks detailed information such as color tunes and structure details present in the original image. 
Although it is not possible to fully recover the details of visual stimuli from noisy brain signals, it is still promising to extract some useful information for visual reconstruction. 
This information can help us generate more realistic reconstructed images. Inspired by this, an alternative branch is introduced to improve the similarity between the generated image and the original visual stimuli. 
This branch includes an autoencoder with an entropy model, which transforms the brain signal into a compact latent feature and then decodes it into a thumbnail image. The purpose of this alternative branch is to provide additional conditional information. 
The overall transform coding process in the SCL is shown in Fig. \ref{fig:Layer3}, and can be described as follows:
\begin{equation}
\begin{aligned}
\boldsymbol{y}_3 &= f_{\text{enc}}^3 (\boldsymbol{x} ; \boldsymbol{\theta}_3) \\
\boldsymbol{\hat{y}_3} &= Q (\boldsymbol{y}_3) \\
\boldsymbol{\hat{z}_3} &= f_{\text{dec}}^3 (\boldsymbol{\hat{y}_3}; \boldsymbol{\phi}_3) .
\end{aligned}
\end{equation}
The encoder $f_{\text{enc}}^3(\cdot ; \boldsymbol{\theta}_3)$ maps the original brain signal $\boldsymbol{x}$ to a latent feature $\boldsymbol{y}_3 \in \mathbb{R}^{C_3}$. The latent feature $\boldsymbol{y}_3$ is rounded and creates the quantized feature $\boldsymbol{\hat{y}}_3 = Q(\boldsymbol{y}_3)$, and the fully factorized entropy model is used to estimate the PDF of the latent feature. The decoder $f_{\text{dec}}^3(\cdot;\boldsymbol{\phi}_3)$ transforms the latent feature $\boldsymbol{\hat{y}}_3$ into a predicted thumbnail $\boldsymbol{\hat{z}}_3 \in \mathbb{R}^{H \times W \times 3}$ with width $W$ and height $H$. 

During training, the rounding operation is replaced by additive uniform noise, and the codec will be optimized to achieve the best trade-off between the bit rate and the distortion of the reconstructed thumbnail according to the loss function:
\begin{equation}
\begin{aligned}
\mathcal L_3 &= R(\boldsymbol{\hat{y}}_3) + \lambda_3 D(\boldsymbol{z}_3, \boldsymbol{\hat{z}}_3) \\ &= \mathbb{E}_{\boldsymbol{x} \sim p_{\boldsymbol{x}}} [{-\log p_{\hat{\boldsymbol{y}_3}}}(\hat{\boldsymbol{y}}_3)]  + \lambda_3 \mathbb{E}_{\boldsymbol{x} \sim p_{\boldsymbol{x}}} [d_1(\boldsymbol{z}_3,\hat{\boldsymbol{z}}_3)],
\end{aligned}
\end{equation}
which aims to minimize the difference between the thumbnail of ground truth visual stimuli and the reconstructed thumbnail transformed from the visual stimuli–evoked brain signal, while the bit rate is constrained. The distortion measure $d_3(\cdot, \cdot)$ is defined as the MSE between the generated thumbnail and the ground truth thumbnail, i.e., 
\begin{equation}
d_3 (\boldsymbol{z}_3, \boldsymbol{\hat{z}}_3) =  \textit{MSE} ( \boldsymbol{z}_3 , \boldsymbol{\hat{z}}_3).
\end{equation}

During inference, the decoder of the OCL and ICL generates the predicted thumbnail and text prompt, which together serve as the condition for a pretrained latent diffusion model \cite{rombach2022high}. This model is then used to generate the image corresponding to the visual stimuli encoded in the brain signal.

\begin{table*}[htbp]

	\centering
	\caption{Evaluation for Brain Signal-based Visual Classification.}
        \resizebox{0.7\textwidth}{!}{
	\begin{tabular}{ccccc} 
        \toprule
		\textbf{Methods} & \textbf{Subject} &  \textbf{Compressed} & \textbf{Bits-Per-Sample} $\downarrow$ & \textbf{Top-1 Accuracy} $\uparrow$  \\ \cline{1-5}
		\multirow{6}{*}{EEGNet \cite{lawhern2018eegnet}} 
          & subj. 1 & \multirow{7}{*}{\XSolidBrush} & \multirow{7}{*}{16} & 19.58\%  \\
		 & subj. 2 &  & & 49.17\% \\
		 & subj. 3 &  & & 48.85\% \\
		 & subj. 4 &  & & 62.61\% \\
		 & subj. 5 &  & & 39.06\% \\
		 & subj. 6 &  & & 46.56\% \\
         % \cline{2-5}
		 & Average ($\pm$ std) &  & & 44.31\% ($\pm$ 13.06\%) \\
        \hline 
        \multirow{6}{*}{EEGChannelNet \cite{palazzo2020decoding}} 
        & subj. 1 & \multirow{7}{*}{\XSolidBrush}  & \multirow{7}{*}{16} & 16.97\% \\ 
        & subj. 2 &  & & 48.80\% \\ 
        & subj. 3 &  & & 49.70\% \\ 
        & subj. 4 &  & & 57.36\% \\ 
        & subj. 5 &  & & 36.36\% \\ 
        & subj. 6 &  & & 41.21\% \\ 
         % \cline{2-5}
		 & Average ($\pm$ std) &  & & 41.73\% ($\pm$ 12.91\%) \\
        \hline 
        \multirow{6}{*}{GRUGate Transformer \cite{tao2021gated}} 
        & subj. 1 & \multirow{7}{*}{\XSolidBrush}   & \multirow{7}{*}{16} & 43.02\% \\ 
        & subj. 2 &   & & 70.52\% \\ 
        & subj. 3 &   & & 63.75\% \\ 
        & subj. 4 &   & & 73.65\% \\ 
        & subj. 5 &   & & 56.25\% \\ 
        & subj. 6 &   & & 64.58\% \\ 
         % \cline{2-5}
		 & Average ($\pm$ std) &  & & 61.96\% ($\pm$ 10.09\%) \\
        \hline 
        \multirow{6}{*}{\textit{EidetiCom} (Proposed) } 
        & subj. 1 & \multirow{7}{*}{\CheckmarkBold} & 0.0168 & 40.00\% \\ 
        & subj. 2 &   & 0.0191 & 60.84\% \\ 
        & subj. 3 &   & 0.0185 & 59.64\% \\ 
        & subj. 4 &   & 0.0174 & 72.37\% \\ 
        & subj. 5 &   & 0.0165 & 55.76\% \\ 
        & subj. 6 &   & 0.0161 & 51.21\% \\ 
         % \cline{2-5}
		 & Average ($\pm$ std) & & 0.0174 ($\pm$ 0.0011)  & 56.64\% ($\pm$ 9.84\%) \\

		% \hline
  %       \hline 
        \bottomrule
	\end{tabular} }
	\label{Brain-to-Label}    
\end{table*}

\section{Experiments}
In this section, we conduct extensive qualitative and quantitative experiments to verify the effectiveness of \textit{EidetiCom}, our innovative cross-modal brain-computer semantic communication paradigm for decoding visual perception.

\subsection{Dataset}
We employ the publicly available visual stimuli–evoked EEG signal dataset ImageNet-EEG \cite{palazzo2020decoding}. The dataset consists of 40 unique object classes, each containing 50 distinct images sourced from the ImageNet dataset \cite{deng2009imagenet}. 
To minimize distraction and optimize user concentration, visual stimuli were presented to 6 participants in a block-based format. Images from each class were displayed consecutively in a single sequence, with each image shown for a duration of 500 ms. 
Additionally, a 10-second black screen was inserted between class blocks to serve as a neutral break period. The EEG signals are acquired using a 128-channel EEG recording device at a sampling rate of 1 kHz, and each sample is represented by the half-precision 16-bit floating point data type.

\textbf{EEG preprocessing.} As standard practice, we conduct the following preprocessing steps on the EEG data. Firstly, the acquired signals are processed by a bandpass filter between 55-95 Hz to avoid extra noise. 
Then, the time window is set to 20-460 ms within the 500 ms time period to avoid interference between different classes. Therefore, the shape of the EEG signal corresponding to visual stimuli after preprocessing is 128 channels $\times$ 440 samples.

\textbf{Image captioning.} The original dataset of visually evoked EEG signals lacks accompanying image captions. To address this gap, we leverage the capabilities of a pretrained image-text multimodal model known as BLIP \cite{li2022blip}. By harnessing the power of BLIP, we are able to seamlessly generate descriptive captions for each corresponding image within the dataset.

\subsection{Implementation Details}
We implement the proposed model using PyTorch and trained on one NVIDIA RTX 3090 GPU. The three layers of the proposed model are trained using Adam optimizer with exponential decay rate $(\beta_1, \beta_2) = (0.9, 0.99)$. The training batch size is set to 16, and the learning rate is set to $1 \times 10^{-4}$. The dataset is divided into training, validation, and test sets that follow the data splitting method as the previous studies \cite{palazzo2020decoding}, with no intersection among them. The model is trained for 150 epochs on the training set, and the model with the best performance on the validation set is selected for evaluation on the test set. The following hyper-parameters of the loss functions are adopted for training: $\lambda_1 = \lambda_3 = 4 \times 10^{4}, \lambda_2 = 40, \alpha_1 = \alpha_2 = 4$. The hidden dimensions of the latent features of three codecs are empirically set to $C_1 = C_2 = 512$, and $C_3 = 2048$ across the experiments.

\subsection{Evaluation Measures}
We evaluate the performance of the proposed brain-computer semantic communication framework for the following three tasks: brain signal-based visual classification, brain-to-caption translation, and brain-to-image generation. 
The metrics are selected to evaluate the compression ratio and the specific task performances. For compression, the bits-per-sample (bps) metric is used to evaluate the average number of bits consumed by the compressed representations. 
This metric is calculated by dividing the total number of bits ($B$) by the number of samples ($N$), expressed as $\textit{bps} = B/N$. For brain signal-based visual classification, the top-1 accuracy is used to evaluate the classification accuracy. 
For brain-to-caption translation, the BiLingual Evaluation Understudy (BLEU) \cite{papineni2002bleu} and Recall-Oriented Understudy for Gisting Evaluation (ROUGE) \cite{lin2004rouge} metrics are used to evaluate the text quality of the reconstructed image captions. 
For brain-to-image generation, the perceptual image quality assessment metrics, such as Inception Score (IS) \cite{salimans2016improved}, Structural Similarity (SSIM) \cite{wang2004image}, Learned Perceptual Image Patch Similarity (LPIPS) \cite{zhang2018unreasonable}, and Deep Image Structure and Texture Similarity (DISTS) \cite{ding2020image} are used to evaluate the quality of the generated images.

\begin{figure*}[htbp]
    \centering
    \includegraphics[width=0.99\linewidth]{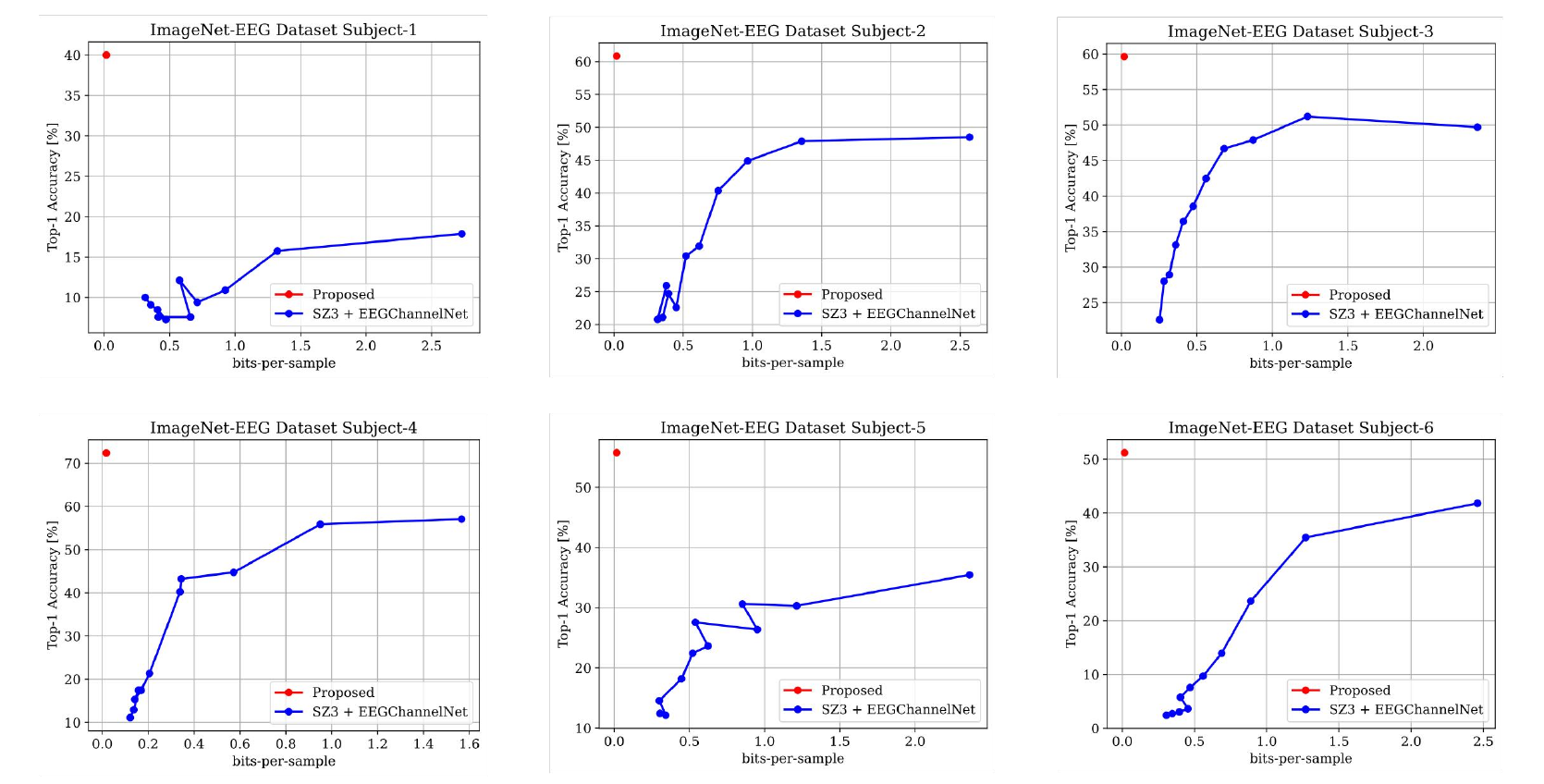}
    \caption{Rate-accuracy performance of brain signal-based visual classification. \textit{SZ3 + EEGChannelNet} denotes the process of compressing EEG signals using the SZ3 algorithm, and subsequently classifying the reconstructed signals using the EEGChannelNet model.}
    \label{fig:rate_acc}
\end{figure*}

\begin{figure}[htbp]
    \centering
    \includegraphics[width=0.99\linewidth]{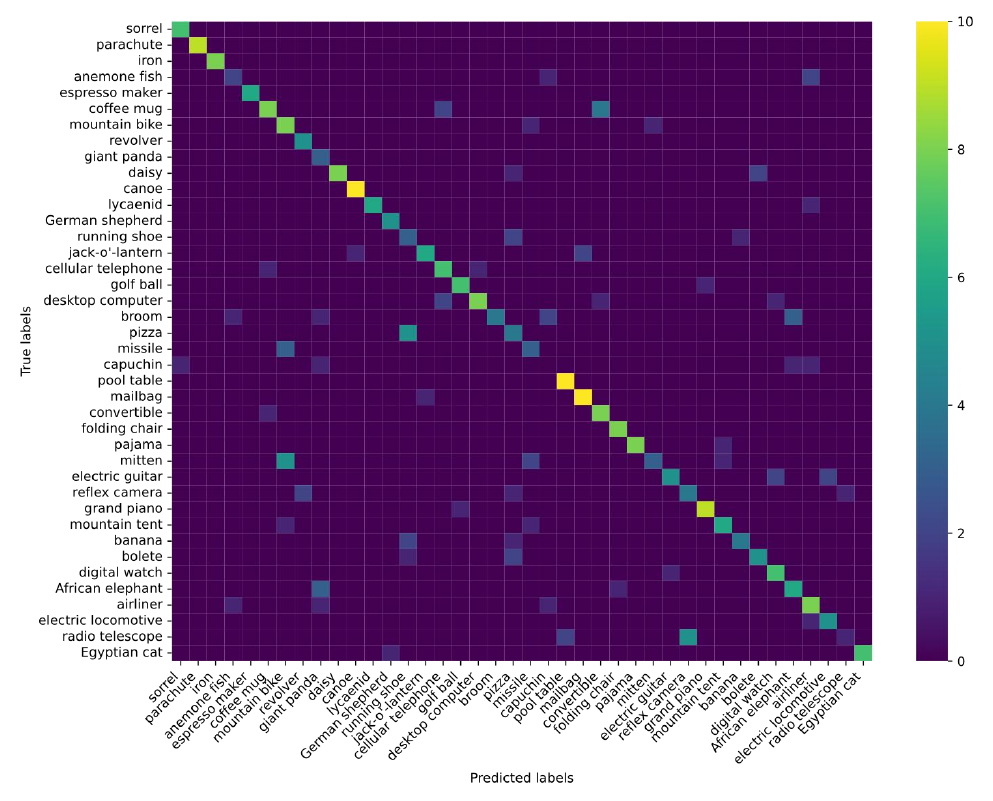}
    \caption{Confusion matrix of brain signal-based visual classification for subject 4. The diagonal elements represent the instances the model predicted correctly. }
    \label{fig:confusion_matrix}
\end{figure}

\subsection{\textit{EidetiCom} for Brain Signal-based Visual Classification}

For brain signal-based visual classification, we exclusively utilize the OCL of our proposed framework. We compare the classification performance of our method with existing state-of-the-art (SOTA) approaches, including EEGNet \cite{lawhern2018eegnet}, EEGChannelNet \cite{palazzo2020decoding}, and GRUGate Transformer \cite{tao2021gated}, over the 6 subjects of the dataset. 
The quantitative results are shown in Table \ref{Brain-to-Label}. It is noteworthy that the compared methods primarily focus on classification accuracy and are not optimized for compression, and we are the first to achieve ultra-low bit rate compression while maintaining comparable classification accuracy to existing methods. 
In contrast, our method achieves an ultra-low bit rate of 0.0174 while maintaining a relatively high top-1 classification performance of 56.64\%. Notably, our method achieves significantly higher classification accuracy compared to EEGNet and EEGChannelNet, and performs comparably to the GRUGate Transformer. 
Furthermore, we compare the rate-accuracy performance of our proposed framework with the combination of the SOTA scientific data compression method SZ3 \cite{liang2022sz3} and EEGChannelNet. As illustrated in Fig. \ref{fig:rate_acc}, our proposed framework demonstrates significantly superior rate-accuracy performance compared to the \textit{SZ3 + EEGChannelNet} combination.

To provide further insights, we present the confusion matrix of the visual classification in Fig. \ref{fig:confusion_matrix}. The majority of categories in the test dataset are accurately classified, with a few exceptions, such as \textit{capuchin} and \textit{radio telescope}, exhibiting slightly lower recognition performance. These results validate that our proposed framework achieves a favorable balance between compression ratio and accuracy for brain signal-based visual classification.

\subsection{\textit{EidetiCom} for Brain-to-Caption Translation}
For brain-to-caption translation, we utilize the OCL to decode the text embedding of labels, while the ICL is conditioned on the text embedding to decode the image captions. To the best of our knowledge, there is no existing work specifically addressing the translation of image captions for visual stimuli-evoked EEG signals. In contrast, our results demonstrate that, even with an extreme compression ratio, our approach remains effective in achieving high-quality translated captions for visual stimuli-evoked EEG signals.

\begin{figure*}[htbp]
    \centering
    \includegraphics[width=0.9\linewidth]{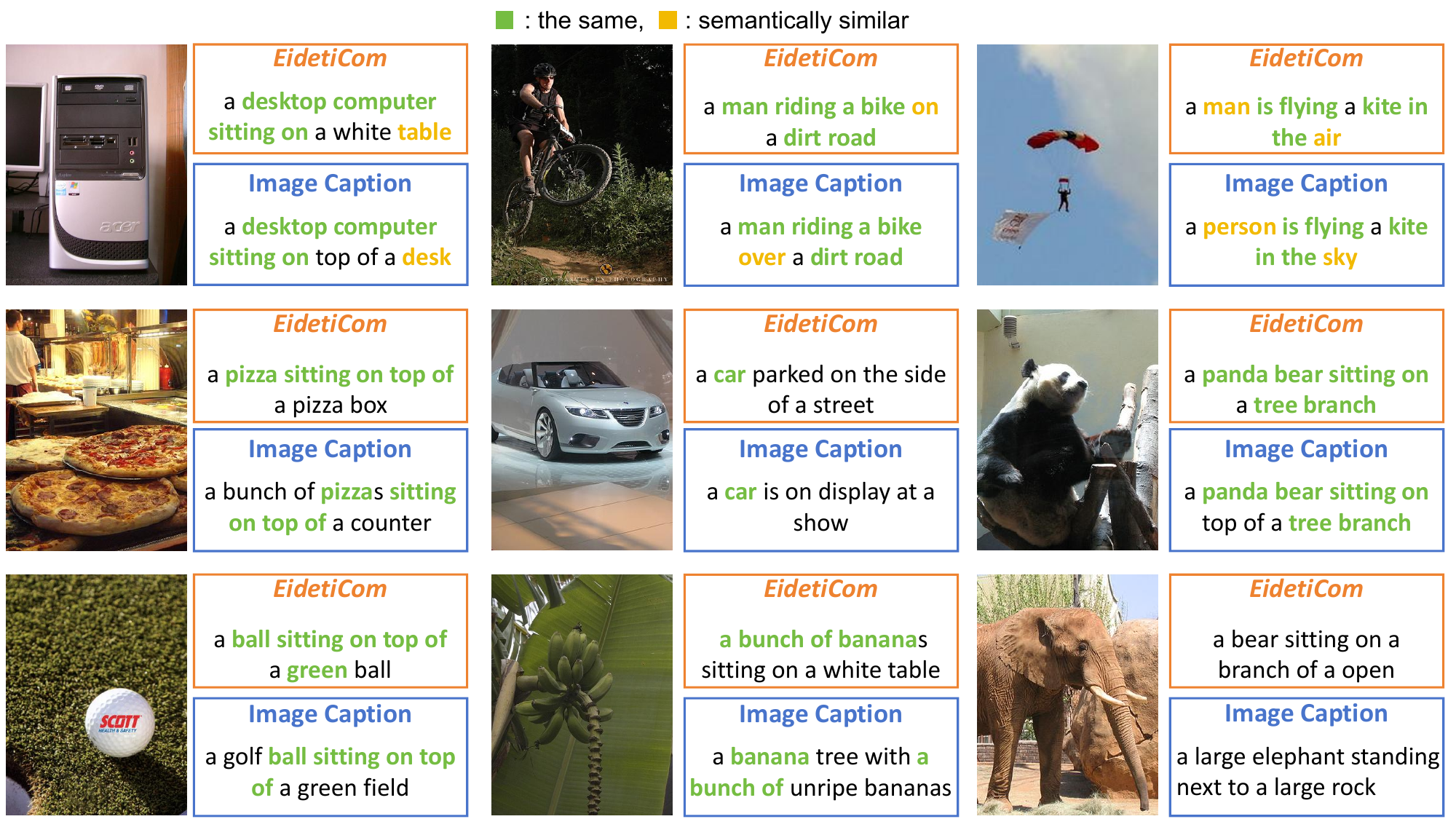}
    \caption{Examples of the translated captions. The image on the left showcases the original image presented to the experimental subjects, while the texts on the right display the ground truth image caption extracted by BLIP and the caption translated from brain signals.}
    \label{fig:caption}
\end{figure*}

We evaluate the bit rate and text quality of the translated captions, as shown in Table \ref{Brain-to-Caption}. The average bits-per-sample of the compressed features is 0.0451. Remarkably, the translated captions achieve notable average BLEU-$n$ scores of 37.79\%, 18.86\%, 8.86\%, and 5.40\% for $n={1,2,3,4}$, respectively. Additionally, the ROUGE-1 scores demonstrate outstanding performance, with a precision score (P) of 42.67\%, a recall score (R) of 41.45\%, and an F1 score (F) of 41.67\%. Fig. \ref{fig:caption} provides examples of the translated captions, compared with the ground truth image captions generated by the pretrained BLIP model. These examples illustrate that, with the assistance of the OCL, the translated image captions effectively convey the category information and even recover additional details related to the visual stimuli.

\begin{table*}[htbp]
	\centering
	\caption{Evaluation for Brain-to-Caption Translation.}
	% \begin{tabular}{c|c|c|cccc|ccc} 
	\resizebox{\textwidth}{!}{
        \begin{tabular}{cccccccccc} 
		% \hline 
  %       \hline
        \toprule 
		\multirow{2}{*}{\textbf{Methods}} & \multirow{2}{*}{\textbf{Subject}} &  \multirow{2}{*}{\textbf{Bits-Per-Sample} $\downarrow$ } &  \multicolumn{4}{c}{\textbf{BLEU-1} $\uparrow$} & \multicolumn{3}{c}{\textbf{ROUGE-1} $\uparrow$ } \\
         & & & \textbf{BLEU-1} & \textbf{BLEU-2} & \textbf{BLEU-3} & \textbf{BLEU-4} & \textbf{P} & \textbf{R} & \textbf{F} \\ \cline{1-10}
		\multirow{7}{*}{\textit{EidetiCom}} 
          & subj. 1 & 0.0461 & 34.36\% & 15.35\% & 6.37\% & 3.83\% & 37.64\% & 38.04\% & 37.51\% \\
		 & subj. 2 & 0.0477 & 39.28\% & 20.86\% & 10.67\% & 6.52\% & 44.18\% & 43.31\% & 43.36\% \\
		 & subj. 3 & 0.0448 & 38.21\% & 19.05\% & 8.16\% & 4.72\% & 42.87\% & 41.65\% & 41.92\% \\
		 & subj. 4 & 0.0423 & 39.57\% & 20.10\% & 9.88\% & 6.15\% & 46.13\% & 43.07\% & 44.15\% \\
		 & subj. 5 & 0.0454 & 38.54\% & 19.72\% & 9.76\% & 6.14\% & 44.27\% & 41.97\% & 42.66\% \\
		 & subj. 6 & 0.0443 & 36.76\% & 18.07\% & 8.30\% & 5.02\% & 40.92\% & 40.64\% & 40.40\% \\
         % \cline{2-10}
		 & Average& 0.0451 & 37.79\% & 18.86\% & 8.86\% & 5.40\% & 42.67\% & 41.45\% & 41.67\%\\
		 & ($\pm$ std) & ($\pm$ 0.0018) & ($\pm$ 1.95\%) & ($\pm$ 1.96\%) & ($\pm$ 1.56\%) & ($\pm$ 1.04\%) & ($\pm$ 3.01\%) & ($\pm$ 1.93\%) & ($\pm$ 2.41\%) \\
        \hline 
        \multirow{7}{*}{\shortstack{\textit{EidetiCom} w/o \\ conditional decoding}} 
        & subj. 1 & 0.0291  & 31.88\% & 13.34\% & 3.76\% & 2.14\% & 34.62\% & 35.32\% & 34.64\% \\ 
        & subj. 2 & 0.0250  & 36.93\% & 17.85\% & 7.70\% & 4.72\% & 40.07\% & 40.79\% & 40.10\% \\ 
        & subj. 3 & 0.0295  & 35.74\% & 16.64\% & 7.01\% & 5.07\% & 39.56\% & 40.27\% & 39.49\% \\ 
        & subj. 4 & 0.0249  & 38.66\% & 20.39\% & 10.69\% & 7.53\% & 43.14\% & 42.62\% & 42.48\% \\ 
        & subj. 5 & 0.0233  & 35.80\% & 17.40\% & 7.47\% & 5.33\% & 38.85\% & 39.13\% & 38.69\% \\ 
        & subj. 6 & 0.0271  & 34.59\% & 15.13\% & 6.11\% & 4.22\% & 38.91\% & 37.36\% & 37.74\% \\ 
         % \cline{2-10}
		& Average & 0.0265 & 35.60\% & 16.79\% & 7.13\% & 4.84\% & 39.19\% & 39.25\% & 38.86\%\\ 
        & ($\pm$ std) & ($\pm$ 0.0025) & ($\pm$ 2.28\%) & ($\pm$ 2.41\%) & ($\pm$ 2.26\%) & ($\pm$ 1.75\%) & ($\pm$ 2.74\%) & ($\pm$ 2.60\%) & ($\pm$ 2.61\%) \\
		% \hline
  %       \hline 
        \bottomrule
	\end{tabular} }
	\label{Brain-to-Caption}    
\end{table*}

\begin{figure*}[htbp]
    \centering
    \includegraphics[width=0.9\linewidth]{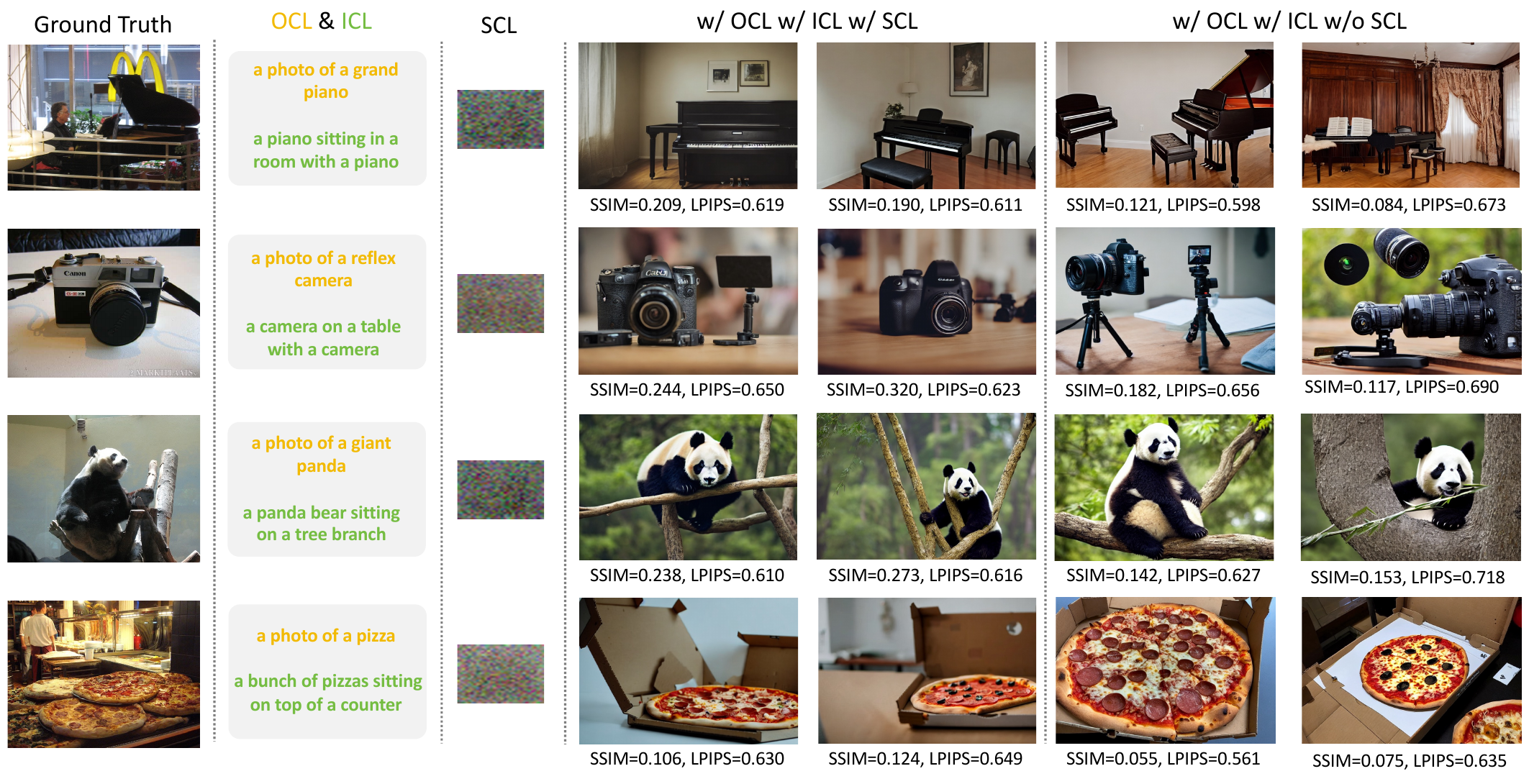}
    \caption{Examples of the generated images. The ground truth visual stimuli, the decoded labels, captions and thumbnails, and the generated image samples with or without the assistance of SCL, are provided.}
    \label{fig:image_samples}
\end{figure*}

\begin{table}[htbp]
	\centering
	\caption{Comparison Experiment for Brain-to-Image Generation.}
	% \begin{tabular}{c|c|c|c|c} 
		% \hline \hline
	\begin{tabular}{ccccc} 
        \toprule
		\textbf{Methods} & \textbf{Compressed}  & \textbf{bps} $\downarrow$ &  \textbf{IS} $\uparrow$ & \textbf{SSIM} $\uparrow$ \\ \cline{1-5}
		Brain2Image \cite{kavasidis2017brain2image} & \XSolidBrush & 16 & 5.01 & - \\
		NeuroVision \cite{khare2022neurovision} & \XSolidBrush & 16 & 5.23 & - \\
		NeuroImagen \cite{lan2023seeing} & \XSolidBrush & 16 & 33.50 & 0.249 \\
		\textit{EidetiCom} & \CheckmarkBold &  0.192 & 28.24 &  0.237 \\
		% \hline \hline
        \bottomrule
	\end{tabular}
	\label{comparison_brain-to-image}    
\end{table}

\subsection{\textit{EidetiCom} for Brain-to-Image Generation}
For brain-to-image generation, we utilize the OCL and ICL to decode the label and image caption, respectively, while the SCL is employed to decode the thumbnail image corresponding to the texture of the visual stimuli. Empirically, we adopt a fusion strategy to integrate both label and caption information. Specifically, when there is an overlap between the nouns in the label and the caption, we use the caption as the text prompt. Conversely, if there is no overlap, we use the label itself as the text prompt. Subsequently, the generative decoder, based on the latent diffusion model, utilizes the text prompts and decoded thumbnails as input to conditionally generate images.

Our approach achieves promising visual reconstruction results for brain-to-image generation, comparing with existing methods. The examples of the generated images, along with the corresponding labels, captions, and thumbnails decoded by the three layers, are illustrated in Fig. \ref{fig:image_samples}. These examples demonstrate that many of the generated images not only capture semantically similar information to the visual stimuli but also exhibit high-quality perceptual similarity. Although recovering low-level information from brain signals is challenging and sometimes impossible due to inherent information loss in the human brain, the SCL enhances the structural similarity scores. Furthermore, as shown in Table \ref{comparison_brain-to-image}, our proposed framework achieves an impressive average IS score of 28.24. Additionally, the average SSIM score is 0.237, while maintaining a low bit rate of 0.192 bits-per-sample. These image quality scores are significantly higher than those of Brain2Image \cite{kavasidis2017brain2image} and NeuroVision \cite{khare2022neurovision}, and comparable to NeuroImagen \cite{lan2023seeing}. It is surprising to find that our approach demonstrates excellent visual reconstruction performance for brain-to-image generation, even with a high compression ratio, while maintaining high-quality perceptual similarity.

\subsection{Impact of Conditional Decoding for Brain-to-caption Translation}
To evaluate the impact of conditional decoding on brain-caption translation, we train the model without conditional decoding and assess its caption generation performance. As shown in Table \ref{Brain-to-Caption}, the translated captions of \textit{EidetiCom} without conditional decoding exhibit a significant decrease in text quality despite consuming a lower bit rate. This can be attributed to the absence of category information provided by OCL, highlighting the importance of conditional decoding.

\begin{table}[htbp]
	\centering
	\caption{Ablation Experiments for Brain-to-Image Generation.}
	% \begin{tabular}{c|c|c|c|c} 
		% \hline \hline
        \setlength\tabcolsep{5pt} 
	\begin{tabular}{ccccccccc} 
        \toprule
		\textbf{\#} & \bf{L1} & \bf{L2} & \bf{L3}  & \textbf{bps} $\downarrow$ &  \textbf{IS} $\uparrow$ & \textbf{SSIM} $\uparrow$ & \textbf{LPIPS} $\downarrow$ & \textbf{DISTS} $\downarrow$ \\ \cline{1-9}
		1 & \XSolidBrush   & \CheckmarkBold & \XSolidBrush   & 0.045 & 26.67 & 0.145 & 0.707 & 0.431 \\
		2 & \XSolidBrush   & \CheckmarkBold & \CheckmarkBold & 0.192 & 24.54 & 0.233 & 0.697 & 0.463 \\
		3 & \CheckmarkBold & \XSolidBrush   & \XSolidBrush   & 0.017 & 28.44 & 0.140 & 0.702 & 0.428 \\
		4 & \CheckmarkBold & \XSolidBrush   & \CheckmarkBold & 0.165 & 27.95 & 0.236 & 0.685 & 0.451 \\
		5 & \CheckmarkBold & \CheckmarkBold   & \XSolidBrush & 0.045 & 28.49 & 0.141 & 0.705 & 0.429 \\
		6 & \CheckmarkBold & \CheckmarkBold   & \CheckmarkBold & 0.192 & 28.24 & 0.237 & 0.685 & 0.451 \\
		% \hline \hline
        \bottomrule
	\end{tabular}
	\label{alation_image_generation}    
\end{table}

\subsection{Impact of Each Layer for Brain-to-Image Generation} 
To investigate the impact of each layer on the brain-to-image generation task, we conduct separate tests by generating images with conditions that included or excluded the outputs of three layers. This allows us to assess the quality of the generated images under different scenarios. The results of the ablation experiments are presented in Table \ref{alation_image_generation}. The findings show that it is feasible to generate high-quality images using label or caption information. However, incorporating SCL notably enhances structural similarity. Moreover, the fusion strategy effectively improves the inception score. It is also noteworthy that the results of the ablation experiments exhibit similarities to those of the NeuroImagen \cite{lan2023seeing}.

\begin{figure}[htbp]
    \centering
    \includegraphics[width=0.7\linewidth]{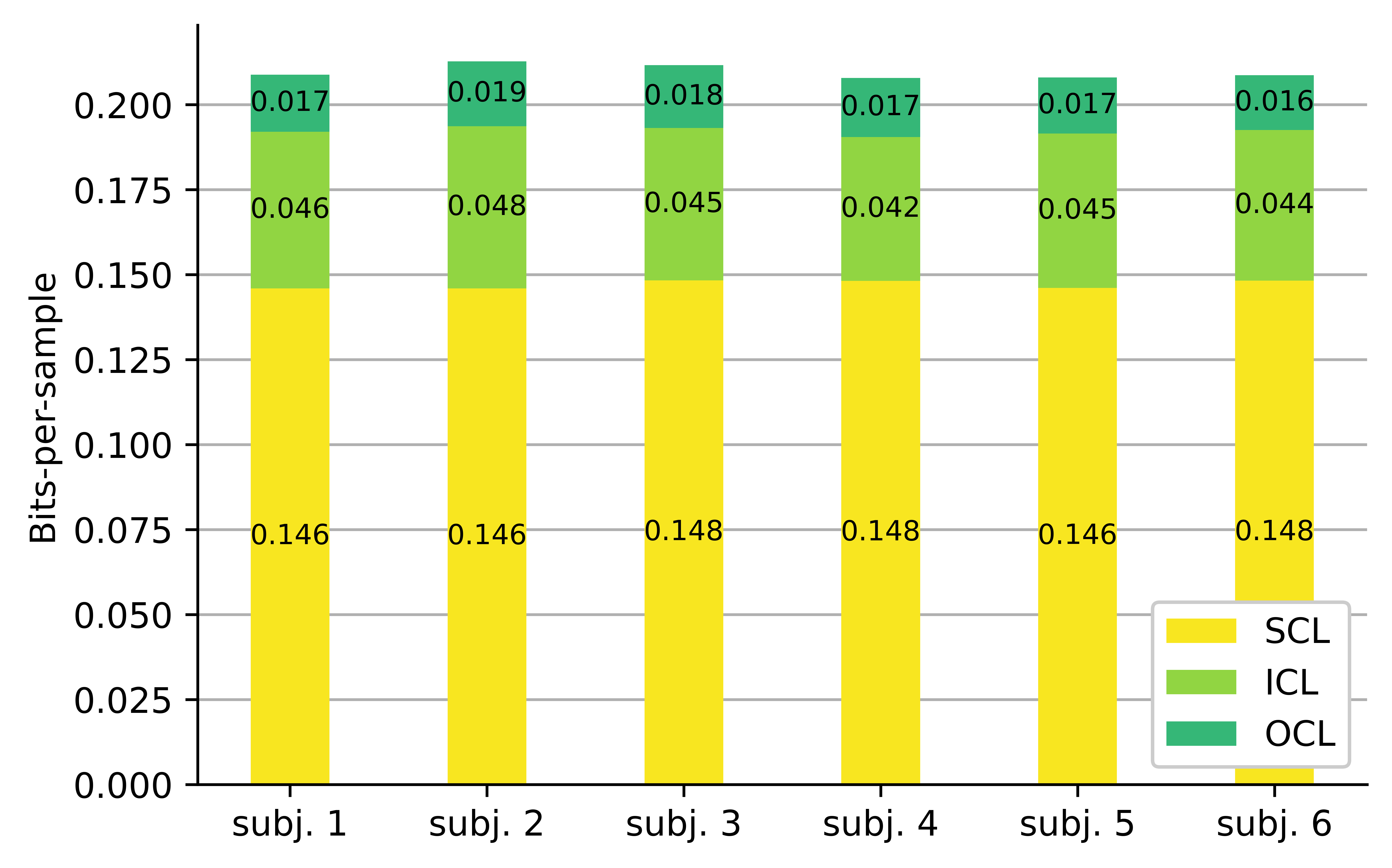}
    \caption{The bit rates consumed by each of the three layers of \textit{EidetiCom}.}
    \label{fig:bit_ablation}
\end{figure}

\subsection{Comparison of the Bit Rates Consumed by Each Layer}
To assess the bit rates consumed by each layer, we conducted separate tests on brain signals from different subjects, focusing on the bit rates consumed by the three layers of semantic compression.As shown in Fig. \ref{fig:bit_ablation}, there is an increase in the bit rates for each layer as the layer ascends. In the bitstream produced by our framework, the category information in OCL necessitates the fewest bits, while the caption information in ICL requires a greater number, and the image information for SCL demands the highest quantity of bits.
Notably, our scalable coding paradigm allows receivers to selectively choose slices of bitstreams based on their unique requirements.

\section{Conclusions}

In our study, we present \textit{EidetiCom}, a cross-modal brain-computer semantic communication paradigm for decoding visual perception. The \textit{EidetiCom} paradigm encapsulates the entire information processing chain, incorporating the information source, semantic encoder, transmission link, semantic decoder, and information destination. 
In this semantic communication framework, the human brain serves as the source, generating visually-evoked brain signals associated with cognition activities. These signals undergo encoding, transmission, and decoding and are ultimately absorbed by the destination, which decodes visual perception for missions.
The proposed framework consists of three hierarchical layers that transform the brain signals into compressed features. The reconstructed semantic features support three semantic-relevant tasks, including brain signal-based visual classification, brain-to-caption translation, and brain-to-image generation. 
The experimental results demonstrate that our proposed framework achieves promising task performances under the condition of limited bit rate for brain-computer communication. 
The proposed paradigm shows potential for enhancing the feasibility of achieving efficient storage and real-time communication for a variety of BCI applications, such as eidetic memory or dream storage, which are sensitive to storage needs, and assistive communication for patients, which is sensitive to bandwidth requirements.

Despite the merits of this study, there are several limitations that highlight areas for future enhancement. First, our current evaluations primarily focus on visually-evoked EEG signals. Exploring evaluations on other modalities of brain signals, such as invasive brain signals, could expand the potential applications of the proposed paradigm. 
Second, the study does not delve into the automatic adaptation to different subjects, which is a significant concern in practical applications. 
Third, the precise mapping between brain recordings and semantic information also remains an open problem, warranting further investigation. 
In conclusion, the development of efficient brain-computer communication frameworks, which take into account both the characteristics of brain signals and practical downstream missions, shows significant potential for reducing the costs associated with storing and transmitting brain signals. This will profoundly impact the real-world application of brain-computer interfaces. Consequently, continued research efforts in this field are imperative.

\bibliographystyle{ieeetr}
\bibliography{Bibliography.bib}

\clearpage

\appendix

\counterwithin{figure}{section}
\counterwithin{table}{section}

\subsection{The Architecture of Object-level Category Layer}
The neural network architecture of the semantic codec of Object-level Category Layer is illustrated in Table \ref{arch_L1}.

\begin{table}[htbp]
	\centering
	\caption{The Architecture of Object-level Category Layer.}
	\begin{tabular}{cccc} 
        \toprule
		\textbf{Module} & \textbf{Input shape}  & \textbf{Output shape} \\ \cline{1-3}
            \textbf{Semantic Encoder} & & & \\
	    Residual conv blocks  & (128, 440) & (,768)  \\
		  Dropout  & (,768) & (,768)  \\
		  Resblock1d  & (,768) & (,1000)  \\
		  Dropout  & (,1000) & (,1000)  \\
		  Resblock1d  & (,1000) & (,512)  \\
            \hline
            \textbf{Semantic Decoder} & & \\
            Resblock1d  & (,512) & (,1000) \\
		Resblock1d  & (,1000) & (,768) \\
		Resblock1d  & (,768) & (,768) \\
            \hline
            \textbf{Entropy Model} & & \\
		Linear  & (,512) & (,512)  \\
		Linear  & (,512) & (,512) \\
		Linear  & (,512) & (,512) \\
		Linear  & (,512) & (,512) \\
        \bottomrule
	\end{tabular}
	\label{arch_L1}    
\end{table}

% \clearpage
\subsection{The Architecture of Image-level Caption Layer}
The neural network architecture of the semantic feature codec of Image-level Caption Layer is illustrated in Table \ref{arch_L2}. In the experimental setting, the CLIP text decoder utilized is derived from the pretrained DeCap model \cite{lidecap}.

\begin{table}[htbp]
	\centering
	\caption{The Architecture of Image-level Caption Layer.}
	\begin{tabular}{cccc} 
        \toprule
		\textbf{Module} & \textbf{Input shape}  & \textbf{Output shape} \\ \cline{1-3}
            \textbf{Semantic Encoder} & & \\
	    Residual conv blocks  & (128, 440) & (,768)  \\
		  Dropout  & (,768) & (,768) \\
		  Resblock1d  & (,768) & (,1000) \\
		  Dropout  & (,1000) & (,1000) \\
		  Resblock1d  & (,1000) & (,512)  \\
            \hline
            \textbf{Semantic Decoder} & &  \\
            Context extracter  & (,512) & (,512) \\
		Feature modulation  & (,512) & (,512) \\
		Resblock1d  & (,512) & (,1000) \\
		Feature modulation  & (,1000) & (,1000) \\
		Resblock1d  & (,1000) & (,768) \\
		Feature modulation  & (,768) & (,768) \\
		Resblock1d  & (,768) & (,512) \\

            \hline
            \textbf{Entropy Model} & &  \\
		Linear  & (,512) & (,512)  \\
		Linear  & (,512) & (,512) \\
		Linear  & (,512) & (,512) \\
		Linear  & (,512) & (,512) \\
        \bottomrule
	\end{tabular}
	\label{arch_L2}    
\end{table}

% \clearpage
\subsection{The Architecture of Stimuli-level Cognition Layer}
The neural network architecture of the semantic feature codec of Stimuli-level Cognition Layer is illustrated in Table \ref{arch_L3}. In the experimental setting, the latent diffusion model utilized is derived from pretrained Stable Diffusion v1-5.

\begin{table}[htbp]
	\centering
	\caption{The Architecture of Stimuli-level Cognition Layer.}
	\begin{tabular}{cccc} 
        \toprule
		\textbf{Module} & \textbf{Input shape}  & \textbf{Output shape}\\ \cline{1-3}
            \textbf{Encoder} & & \\
	    Linear  & (128, 440) & (,4096)  \\
		  Dropout  & (,4096) & (,4096)  \\
		  Resblock1d  & (,4096) & (,3072)  \\
		  Dropout  & (,3072) & (,3072)  \\
		  Resblock1d  & (,3072) & (,2048)  \\
            \hline
            \textbf{Decoder} & & \\
            Resblock1d  & (,2048) & (,3072) \\
		Resblock1d  & (,3072) & (,4096) \\
		Resblock1d  & (,4096) & (,3072) \\
            \hline
            \textbf{Entropy Model} & & \\
		Linear  & (,2048) & (,2048)  \\
		Linear  & (,2048) & (,2048) \\
		Linear  & (,2048) & (,2048) \\
		Linear  & (,2048) & (,2048) \\
        \bottomrule
	\end{tabular}
	\label{arch_L3}    
\end{table}

% \clearpage
\subsection{Examples of the translated image captions}
The examples of the translated image captions are depicted in Table \ref{caption_examples}. It effectively shows that, despite some loss of semantic information, many meanings in image captions remain translatable to varying extents.

\begin{table*}[htbp]
	\centering
	\caption{Examples of the translated image captions.}
        \resizebox{0.8\linewidth}{!}{
	\begin{tabular}{cccc} 
        \toprule
		\textbf{Ground Truth (BLIP)} & \textbf{Translated} \\ \cline{1-2}

a group of people riding bikes down a dirt road & a pair of gloves sitting on a small table  \\ 
a little girl standing next to a window with cupcakes on it & a couple of people standing by a small bed on a  \\ 
a man in a red sweater playing a piano & a piano in a black and white table  \\ 
a couple of brown horses standing inside of a barn & a brown horse standing on a large green hill  \\ 
a close up of a golf ball in the grass & a ball sitting on top of a green ball  \\ 
a pool table with a pool ball on it & a table that has a ball and a green top  \\ 
an american airlines plane flying in the sky & a brown elephant sitting on the front of its head  \\ 
a pool table in a living room with a fireplace in the background & a green table with a ball inside of it  \\ 
a black and white cat walking across a dirt road & a cat sitting on a gray cat on top of a bed  \\ 
a person's hand with a tattoo on it & a train sitting on a piece of a white metal  \\ 
a large white rocket sitting on top of a lush green field & a plane sitting in the middle of a line of some sort  \\ 
a bunch of yellow flowers that are in the grass & a piece of flowers sitting on a ground with a white head  \\ 
a woman standing in front of a display of brooms & a single bear sitting on a white surface  \\ 
an airplane is flying over a field of crops & a plane sitting on the ground with a very high  \\ 
a close up of a horse with a black background & a horse standing on a very long brown field  \\ 
a grand piano sitting on a hard wood floor & a piano in a black and white table  \\ 
a young horse running in a fenced in area & a horse standing on a brown horse on a field  \\ 
a close up of a coffee cup on a table & a car sitting on a table with a hand  \\ 
a man standing next to a tent on top of a mountain & a man riding a bike on a small ground  \\ 
a brown and white butterfly sitting on top of a white flower & a airplane sitting on a small line of some green  \\ 
a man and a woman standing in front of a fruit stand & a pair of bananas on a white table  \\ 
a miniature man standing next to a pool ball & a green table that has two balls in a room  \\ 
a glass bowl with a wooden spoon on a table & a ball of a ball sitting on top of a green surface  \\ 
a bottle of ink sitting next to a wooden bag & a close up of a white chair on a table  \\ 
a couple of flowers that are in a vase & a piece of flowers sitting on a ground with a brown head  \\ 
a man riding a mountain bike on a rocky trail & a a plane on a ground with a bunch of water  \\ 
a close up of a coffee machine with a cup of coffee & a coffee cup sitting on top of a table  \\ 
a close up of a camera on a white surface & a picture of a vehicle with a big camera on top of it  \\ 
a couple of dogs that are running in the grass & a dog and a dog laying on the ground  \\ 
a young boy sitting in a chair with a guitar & a close up of a clock sitting on a bed  \\ 
a basket full of golf balls on a putting green & a picture of a piano on a brown table  \\ 
a bunch of white flowers with yellow centers & a pair of bananas sitting on a white surface  \\ 
a group of people holding a rainbow colored kite & a person on a parachute flying in the air  \\ 
a gun in a case on a table & a camera that is on a black table  \\ 
a computer monitor sitting on top of a wooden desk & a close up of a phone with a sitting on a table  \\ 
a red coffee cup sitting on top of a wooden table & a car sitting on the side of a street  \\ 
a lit up pumpkin sitting on a porch & a apple sitting on a black bag on the ground  \\ 
a close up of a shoe on a table & a pair of bananas that are on a white table  \\ 
a white and black butterfly sitting on top of a plant & a small brown bird on a piece of leaves on a green plant  \\ 
a dark room with three computer monitors and a chair & a close up of a computer sitting on a desk  \\ 
a silhouette of a man playing a guitar & a train on a train tracks next to a blue building  \\ 
a close up of a mushroom on a blue surface & this is a white banana in the middle of a forest  \\ 
a close up of a mushroom on the ground & a closeup of a branch with a bunch of white  \\ 
a carved pumpkin with a woman's face on it & a orange on a black bag on a table  \\ 
a person holding a coffee mug with a picture of a hand on it & a car sitting on a road next to a cup  \\ 
a brown horse standing on top of a lush green field & a horse standing in a large brown horse on some ground  \\ 
a white iron sitting on top of a wooden table & a close up of a white chair on a table  \\ 
a brown dog with a blue leash on it's neck & a dog laying in the grass with a dog  \\ 
a white plate topped with sliced mushrooms on top of a table & this is a white banana sitting on the ground  \\ 
a black and white cat laying inside of a brown paper bag & a apple sitting on a black bag on the ground  \\ 
a painting of a man and his dog in a canoe & a purse sitting on a black top  \\ 
a pink purse sitting on top of a bed & a apple sitting on a black surface  \\ 
two people are in the air with parachutes & a person on a parachute flying in the air  \\ 
        \bottomrule
	\end{tabular}
        }
	\label{caption_examples}    
\end{table*}

% \clearpage
\subsection{Examples of the generated images}
\begin{figure*}[h]
\begin{adjustwidth}{2.2cm}{0.2cm} 
    \centering
    \includegraphics[width=0.7\linewidth]{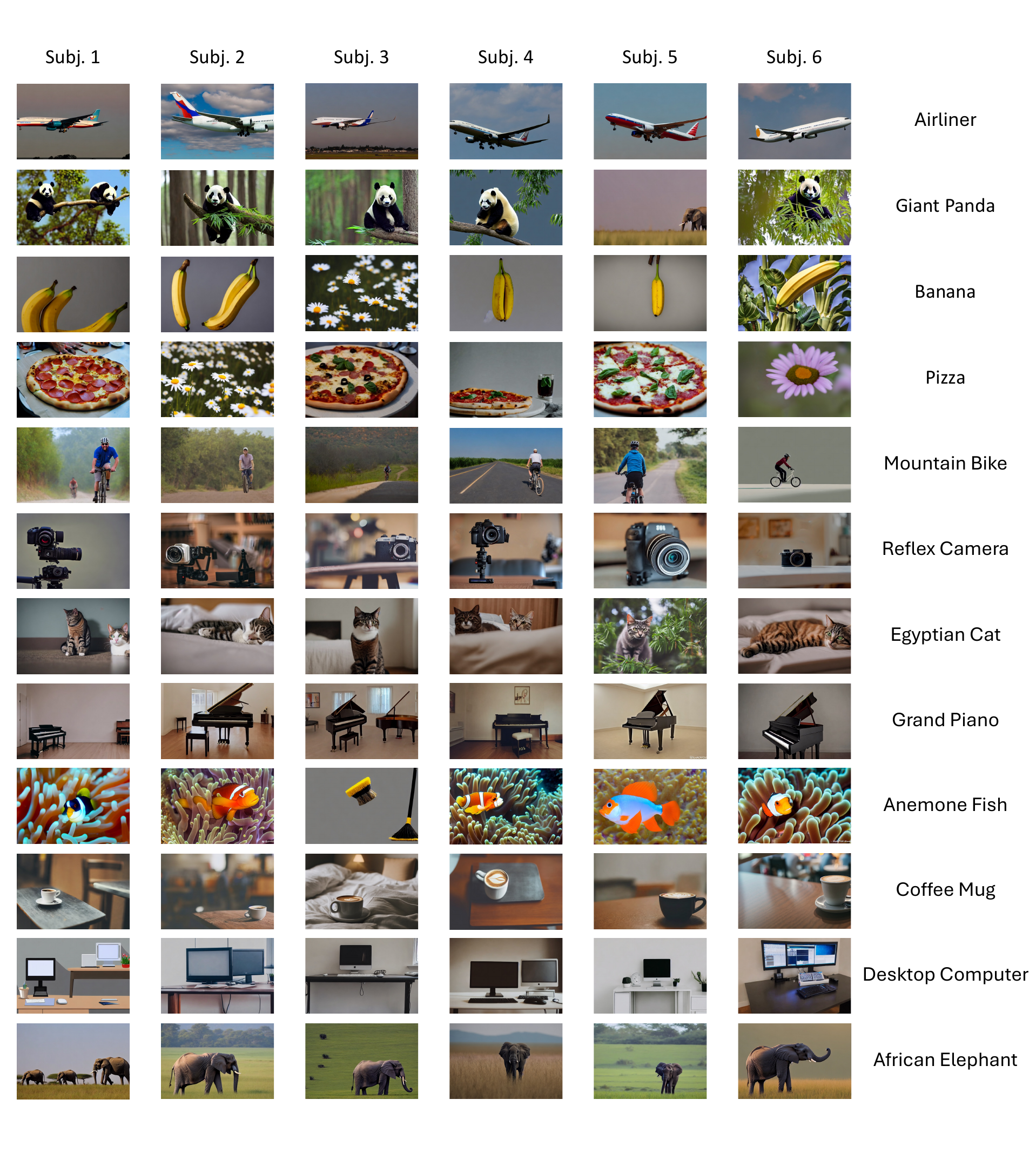}
    \caption{Examples of the generated images.}
    \label{fig:image_examples}
\end{adjustwidth}
\end{figure*}

The examples of the generated images are shown in Fig. \ref{fig:image_examples}. The results effectively demonstrate that, despite encountering challenges such as misclassification, a significant portion of the generated images retain the capability to convey relevant visual information through the brain-computer semantic communication paradigm.

% \clearpage
\subsection{Impact of Generative Models}
\begin{figure*}[h]
\begin{adjustwidth}{0cm}{0.7cm} 
    \centering
    \includegraphics[width=0.7\linewidth]{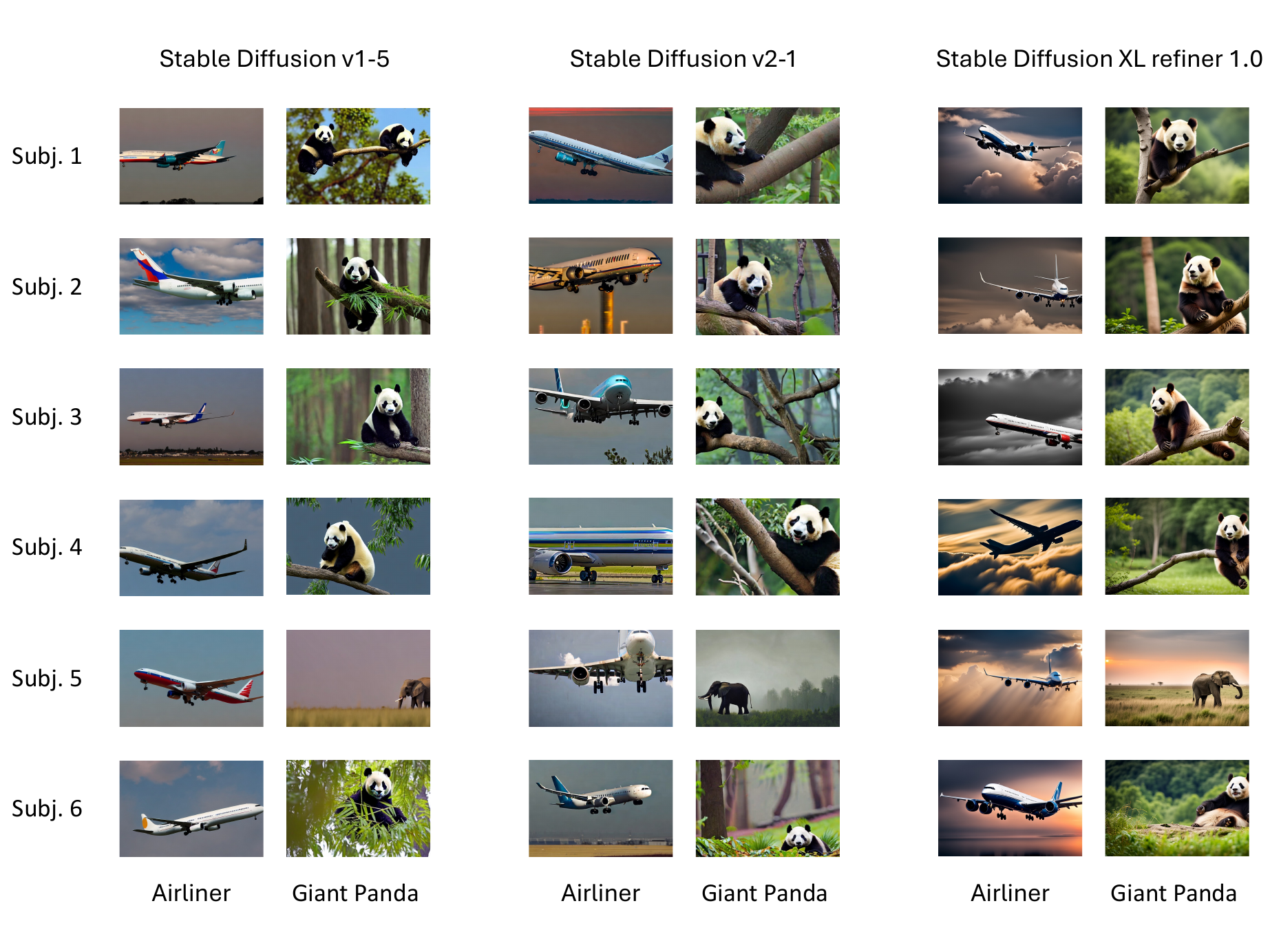}
    \caption{Subjective comparison of the images generated by different generative models.}
    \label{fig:compare_gen_model}
\end{adjustwidth}
\end{figure*}

Due to its modular design, the framework proposed in this study enables the seamless integration of more advanced generative models. In our experiments, we employ a pretrained Stable Diffusion v1-5 model as the latent diffusion model. With the swift evolution of generative model technology, numerous more sophisticated models, such as Stable Diffusion XL, have emerged. We present the experimental results of applying different generative models to our proposed framework in Table \ref{impact_of_gen_model} and Fig. \ref{fig:compare_gen_model}.

\begin{table*}[htbp]
	\centering
	\caption{Impact of Generative Models}
        \setlength\tabcolsep{5pt} 
	\begin{tabular}{cccccc} 
        \toprule
		\textbf{Generative Model} & \textbf{bps} $\downarrow$ &  \textbf{IS} $\uparrow$ & \textbf{SSIM} $\uparrow$ & \textbf{LPIPS} $\downarrow$ & \textbf{DISTS} $\downarrow$ \\ \cline{1-6}
		Stable Diffusion v1-5 & 0.192 & 24.54 & 0.233 & 0.697 & 0.463 \\
		Stable Diffusion v2-1 & 0.192 & 24.53 & 0.235 & 0.702 & 0.469 \\
		Stable Diffusion XL refiner 1.0 & 0.192 & 24.57 & 0.197 & 0.706 & 0.443 \\
  
        \bottomrule
	\end{tabular}
	\label{impact_of_gen_model}    
\end{table*}

\vfill

\end{document}